# Phase Diagrams of Single Layer Two-Dimensional Transition Metal Dichalcogenides: Landau Theory

Anna N. Morozovska[1,*], Eugene A. Eliseev[2], Kevin D. Stubbs[3], Rama Vasudevan[3], Yunseok Kim[4], and Sergei V. Kalinin[3,†]

[1]*Institute of Physics, National Academy of Sciences of Ukraine,*
*46, pr. Nauky, 03028 Kyiv, Ukraine*

[2] *Institute for Problems of Materials Science, National Academy of Sciences of Ukraine,*
*Krjijanovskogo 3, 03142 Kyiv, Ukraine*

[3]*The Center for Nanophase Materials Sciences, Oak Ridge National Laboratory,*
*Oak Ridge, TN 37922*

[4] *School of Advanced Materials Science and Engineering, Sungkyunkwan University (SKKU), Suwon 16419, Republic of Korea*

**Abstract**

Single layer (SL) two-dimensional transition metal dichalcogenides (TMDs), such as $MoS_2$, $ReS_2$, $WSe_2$, and $MoTe_2$ have now become the focus of intensive fundamental and applied researches due to their intriguing and tunable physical properties. These materials exhibit a broad range of structural phases that can be induced via elastic strain, chemical doping, and electrostatic field effect. These transitions in turn can open and close the band gap of SL-TMDs, leading to metal-insulator transitions, and lead to emergence of more complex quantum phenomena. These considerations necessitate detailed understanding of the mesoscopic mechanisms of these structural phase transitions. Here we develop the Landau-type thermodynamic description of SL-TMDs on example of SL-$(MoS_2)_{1-x}$-$(ReS_2)_x$ system and analyze the free energy surfaces, phase diagrams, and order parameter behavior. Our results predict the existence of multiple structural phases with 2-, 6- and 12-fold degenerated energy minima for in-plane and out of plane order parameters. This analysis suggests that out-of-plane ferroelectricity can exist in many of these phases, with the switchable polarization being proportional to the out-of-plane order parameter. We further predict that the domain walls in SL-$(MoS_2)_{1-x}$-$(ReS_2)_x$ should become conductive above a certain strain threshold.

---

[*] Corresponding author 1: anna.n.morozovska@gmail.com

[†] Corresponding author 2: sergei2@ornl.gov

# I. INTRODUCTION

Intriguing tunable physical properties of single layer (**SL**) two-dimensional (**2D**) transition metal dichalcogenides (**TMDs**), such as $MoS_2$, $ReS_2$, $WSe_2$, and $MoTe_2$ have made them an object of intensive fundamental and applied research [1, 2, 3, 4]. A significant volume of fundamental studies is devoted to the symmetry and domain structure [5, 6], band structure, elastic and electronic properties of SL-TMDs [7, 8, 9, 10, 11, 12], and demonstrate various possibilities to tune the properties in a wide range due to mechanical strain [4, 5, 6, 7], chemical doping, and electrostatic field effect [12]. SL-TMDs, such as $MoS_2$, $ReS_2$ and $WSe_2$, unlike gapless graphene sheets and indirect gap bulk TMDs, are direct gap semiconductors, and therefore can be widely used in optoelectronics devices [13, 14, 15, 16]. Similarly, a number of applications of hybrid structures of SL-TMDs with bilayer graphene [17] (possibility to tune electronic structure of $MoS_2$), and as ferroelectrics [18] (ultra-fast, non-volatile multilevel memory devices with non-destructive low-power readout) have emerged.

These materials show remarkable tunability of phase transitions via external stimuli. For example, application of mechanical strain can switch SL-TMDs between several phases, corresponding to the irreducible representations of its 2D symmetry and having different electronic properties [5, 6]. The associated changes in the band gap of SL-TMDs can lead to metallic transition [4], allowing tuning electronic properties of SL-TMDs from semiconducting to metallic [4, 6]. These considerations necessitate developing detailed composition-dependent structural phase diagrams of these materials, and corresponding Landau theory to predict the dopant, strain, and field effects. Here, we construct and explore the Landau-type thermodynamic description of SL-$MoS_2$-$ReS_2$ system.

To construct the compositional phase diagram of SL-$(MoS_2)_{1-x}$-$(ReS_2)_x$ within Landau approach, one has to reproduce the symmetry properties of the respective end members, e.g. $MoS_2$ (x=0) and $ReS_2$ (x=1). The point group of unstrained $MoS_2$ unit cell is 6/mmm [19] (trigonal prismatic $D_{3h}$), with the "trigonal prismatic" semiconducting H phase stable under normal conditions. Rigorously speaking, the $\bar{6}m2$ point group is reduced to $6mm$ point group for the infinite and flat SL-$MoS_2$, because all "out-of-plane" symmetry operations moving atoms away from the SL plane disappear [20]. The semiconducting H phase are connected with metallic octahedral T or semi-metallic T' phases via displacive structural transformations permitting rapid and possibly reversible switching between the structures in SL-$MoS_2$ [21]. On the other hand, H phase can be transformed to the T phase by Re atom doping [22], though T phase is metastable without the doping or other specific treatment. Under the specific treatment H phase can also be transformed to the distorted octahedral phase with clusterization of metal atoms into zigzag chains in "Z" phase of SL-$MoS_2$ [23].

Bulk ReS$_2$ exists in the diamond-shape (DS) phase with triclinic symmetry, where the neighboring Re clusters are linked along the b[010] axis to form Re DS-chains [24]. The point group of unstrained bulk ReS$_2$ unit cell is triclinic $C_i$ [25, 26]; however, higher symmetry polymorphs such as 3m [27] are possible. For the SL-ReS$_2$, as 2D system, the point group remains unchanged (as containing no "out-of-plane" symmetry operations) and stable under normal conditions. Lin et al [28] showed here that SL-ReS$_2$ is a stable n-type semiconductor which exhibits an order of magnitude difference in anisotropic conduction, unlike most 2D materials. Note that most MoS$_2$ and ReS$_2$ structures cannot coexist without defects and broken bonds [4], suggesting the potential for intriguing atomic level mechanisms.

Previously Berry et al. [5, 6] has proposed the continuum Landau-type thermodynamic potential for 2D-TMDs with chemical formulae MX$_2$ (M – metal, X – chalcogenide). In Berry description [5, 6], the free energy density includes Landau part ($f_L$) and elastic energy ($f_{EL}$) that in turn consists of the intrinsic strain and extrinsic bending energy. However, Berry et al [5, 6] considered only two phases for TMDs, namely H phase (with zero order parameter, $\eta=0$) and T' phase (with non-zero order parameter, $\eta=+1$). The T phase was ignored as unstable. At the same time, the full free energy functional for SL TMDs should allow the continuous strain-induced and composition-induced transformations between four (or even five) symmetry phases, namely the "host MoS$_2$" H, (meta)stable T and stable T' phases, and "guest ReS$_2$" DS phase, and possible fifth Z phase. In the case of SL-(MoS$_2$)$_{1-x}$-(ReS$_2$)$_x$ the Berry $\eta=0$ in the T phase, and hence an extra energy term describing the transition between H and T phases should be included. Here we construct the thermodynamic potential of SL-(MoS$_2$)$_{1-x}$-(ReS$_2$)$_x$ system allowing for all possible phases.

The paper is organized as following. **Section II** contains the derivation of the Landau-type free energy functional for SL-(MoS$_2$)$_{1-x}$-(ReS$_2$)$_x$ and equations of state for vector order parameter. Free energy functional of MoS$_2$ is analyzed in **Section III.** Free energy, phase diagrams and order parameter behavior of SL-(MoS$_2$)$_{1-x}$-(ReS$_2$)$_x$ are studied in **Section IV.** Obtained results are discussed and summarized in **Section V** and **VI**, respectively**. Supplementary Materials** contain the calculation details listed in **Appendixes A** and **B**.

## II. LANDAU-TYPE FREE ENERGY FUNCTIONAL

To modify Berry et al. [5, 6] approach for multi-phase transformations, we introduce a vector order parameter, $\eta$, being an analog of spontaneous strain tensor or polarization/magnetization vectors in ferroics. The nature of this order parameter can be established from the statistical normal mode studies of experimental high resolution scanning transmission electron microscopy data on SL-(MoS$_2$)$_{1-x}$-(ReS$_2$)$_x$, system [29]. This mode represents the dominant symmetry breaking mode for all studied compositions. Below we assume that this mode is associated with vector order parameter $\boldsymbol{\eta}$. The continuum approach allows us plot the phase state of the order parameter as a function of Mo/Re ratio x.

The Landau-type part of the free energy density for $(MoS_2)_{1-x}$-$(ReS_2)_x$ system is expanded as

$$f_L[\boldsymbol{\eta}(\vec{r})] = \alpha_{ij}\eta_i\eta_j + \beta_{ijkl}\eta_i\eta_j\eta_k\eta_l + \gamma_{ijklmn}\eta_i\eta_j\eta_k\eta_l\eta_m\eta_n \qquad (1)$$

where $x$ is the fraction of Mo/Re atoms. Note that subscripts $i,j,k,l=x,y,z$ for a 3D order parameter $\boldsymbol{\eta} = (\eta_1, \eta_2, \eta_3)$ corresponding to bulk or SL-$(MoS_2)_{1-x}$/$(ReS_2)_x$ ($x>0$), and $i,j,k,l=x,y$ for a 2D order parameter $\boldsymbol{\eta} = (\eta_1, \eta_2, 0)$ inherent to SL-$MoS_2$. The orientation of $\boldsymbol{\eta}$ with respect to $(MoS_2)_{1-x}$-$(ReS_2)_x$ atomic structure is shown in **Fig. 1**. Following the Landau analysis of traditional perovskite systems, we assume that the tensorial expansion coefficients $\alpha_{ij}$, $\beta_{ijkl}$ and $\gamma_{ijklmn}$ in Eq.(1) can be composition $x$-dependent.

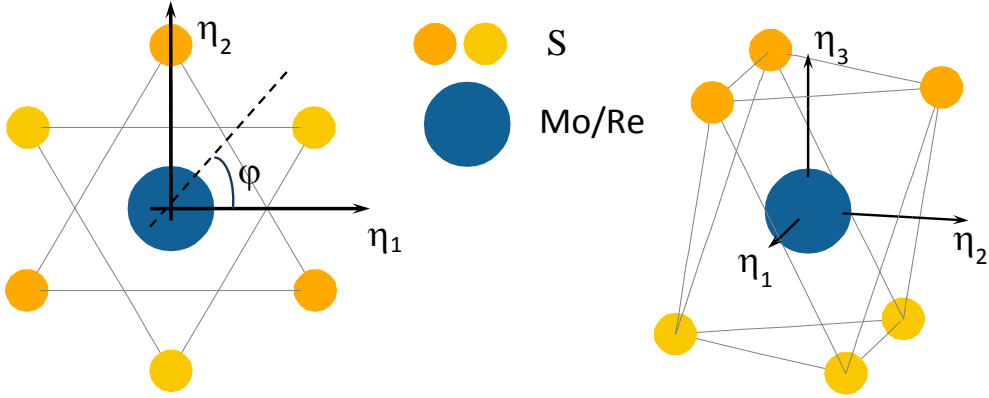

**FIGURE 1.** Orientation of the order parameter components $\{\eta_1, \eta_2, \eta_3\}$ in SL-$(MoS_2)_{1-x}$-$(ReS_2)_x$. The SL lies in the $\{\eta_1, \eta_2\}$ plane. Metal (Mo or Re) and upper/lower chalcogen (S) atoms are shown in gray-blue and orange/yellow colors, respectively.

The general form of the elastic free energy is chosen to be compatible with all domain configurations [5, 6], namely:

$$f_{EL}[\boldsymbol{\eta}(\vec{r}), u_{kl}] = \frac{c_{ijkl}}{2} u_{ij} u_{kl} + \left( Q_{ijkl}\eta_k\eta_l + R_{ijklmn}\eta_k\eta_l\eta_m\eta_n \right) u_{ij}, \qquad (2)$$

where $c_{ijkl}$ are the components of elastic stiffness tensor, $u_{kl}$ are the components of elastic strain tensor, and $Q_{ijkl}$ and $R_{ijklmn}$ are the components of striction tensors, respectively.

Since most possible crystallographic MoS$_2$-ReS$_2$ structures cannot coexist without defects and broken bonds [4], the total free energy $f[\mathbf{\eta}]$ density should also include the gradient and flexoelectric terms energy to describe associated local deformations. Finally, $f[\mathbf{\eta}]$ acquires the form:

$$f[\mathbf{\eta}] = f_L[\mathbf{\eta}(\vec{r})] + f_{EL}[\mathbf{\eta}(\vec{r})] + \frac{g_{ijkl}}{2}\frac{\partial \eta_i}{\partial x_j}\frac{\partial \eta_k}{\partial x_l} + \frac{f_{ijkl}}{2}\left(\frac{\partial \eta_i}{\partial x_j}u_{kl} - \eta_i\frac{\partial u_{kl}}{\partial x_j}\right). \quad (3)$$

Depending on the system dimensionality (3D or 2D) the free energy functional reads:

$$F_{3D}[\mathbf{\eta}] = \int_V d^3r f[\mathbf{\eta}(\vec{r})], \qquad F_{2D}[\mathbf{\eta}] = \int_S d^2r f[\mathbf{\eta}(\vec{r})] \quad (4)$$

As the next step, we establish the symmetry of tensors in Eq. (3). In Landau theory, the free energy is invariant with respect to all point group symmetry transformations of the "parent" or "aristo" phase. Hence, for 3D bulk (MoS$_2$)$_{1-x}$/(ReS$_2$)$_x$, we should use $\bar{6}m2$ (for x=0) and $3m$ (for x=1) point groups, respectively. For a 2D SL-(MoS$_2$)$_{1-x}$/(ReS$_2$)$_x$, we should consider *6mm* and *3m* point groups, respectively. Since *6mm* is a higher symmetry, we use this symmetry group to describe SL-MoS$_2$ aristo-phase (x=0), and pose that doping with Re leads to symmetry lowering transitions to 3m group for SL-ReS$_2$ (x=1). The possible domain configurations can then be obtained in lower symmetry phases in accordance with Curie principle.

Direct application of aristo-phase point group symmetry operations to the tensors $\alpha_{ij}$, $\beta_{ijkl}$, $\gamma_{ijklmn}$, $g_{ijkl}$, $Q_{ijkl}$ and $R_{ijklmn}$ (shortly, the "direct method" [30]) along with existing index-permutation symmetry, leads to the system of linear algebraic equations for each tensor, e.g. $\alpha_{ij} = C_{ii'}C_{jj'}\alpha_{i'j'}$ (see **Appendix A**). The first lines in these equations include point group symmetry, where matrix elements $C_{ii'}$ are $\bar{6}m2$ (for a bulk), *6mm* or *3m* (for a SL) group symmetry operations in the matrix form [31], and the second line reflects the index-permutation symmetry (if any exist). Solution of these equations gives the detailed structure of the tensors, including the number of nonzero and independent elements, as listed in **Appendix A.**

Setting the strain and spatial gradients equal to zero for the considered free SL, the free energy density acquires the form

$$f_L[\mathbf{\eta}(\vec{r})] = f_L^{6mm}[\mathbf{\eta}(\vec{r})] + \Delta f_L^{\bar{3}m}[\mathbf{\eta}(\vec{r})]. \quad (5)$$

The parts with 6mm symmetry and specific for $\bar{3}m$ symmetry are:

$$\begin{aligned}f_L^{6mm}[\mathbf{\eta}(\vec{r})] = &\alpha_{11}\eta_\perp^2 + \alpha_{33}\eta_3^2 + \beta_{11}\eta_\perp^4 + \beta_{13}\eta_\perp^2\eta_3^2 + \beta_{33}\eta_3^4 + \gamma_{113}\eta_\perp^4\eta_3^2 + \gamma_{133}\eta_\perp^2\eta_3^4 \\ &+ \gamma_{111}\eta_1^2(\eta_1^2 - 3\eta_2^2)^2 + \gamma_{222}\eta_2^2(3\eta_1^2 - \eta_2^2)^2 + \gamma_{333}\eta_3^6\end{aligned} \quad (6)$$

$$\Delta f_L^{\bar{3}m}[\mathbf{\eta}(\vec{r})] = \eta_1\eta_3(\eta_1^2 - 3\eta_2^2)(\beta_{15} + \gamma_{115}\eta_\perp^2 + \gamma_{135}\eta_3^2). \tag{7}$$

where $\eta_\perp = \sqrt{\eta_1^2 + \eta_2^2}$ is the absolute value of the in-plane order parameter. The 4$^{th}$ and 6$^{th}$ power terms in Eqs.(6)-(7) break the transverse in-plane isotropy of the free energy density (5) with respect to the orientation of order parameter components $\eta_i$. Not negativity of coefficients $\gamma_{111} \geq 0$, $\gamma_{222} \geq 0$ and $\gamma_{333} \geq 0$ is required for thermodynamic stability.

Eq.(5) yields three coupled equations of state for the components of the order parameter $\mathbf{\eta} = (\eta_1, \eta_2, \eta_3)$, listed in **Appendix B**. Substituting $\eta_1 = \eta_\perp \cos(\varphi)$ and $\eta_2 = \eta_\perp \sin(\varphi)$ in Eqs.(6)-(7), we obtain:

$$\begin{aligned}f_L[\mathbf{\eta}(\vec{r})] &= \alpha_{11}\eta_\perp^2 + \alpha_{33}\eta_3^2 + \beta_{11}\eta_\perp^4 + \beta_{13}\eta_\perp^2\eta_3^2 + \beta_{33}\eta_3^4 + \gamma_{113}\eta_\perp^4\eta_3^2 + \gamma_{133}\eta_\perp^2\eta_3^4 + \gamma_{333}\eta_3^6 \\ &+ (\gamma_{111}\cos^2(3\varphi) + \gamma_{222}\sin^2(3\varphi))\eta_\perp^6 + (\beta_{15} + \gamma_{115}\eta_\perp^2 + \gamma_{135}\eta_3^2)\eta_\perp^3\eta_3\cos(3\varphi)\end{aligned} \tag{8}$$

Equations (8) cannot be solved analytically for arbitrary values of material parameters, but several important analytical solutions valid for particular cases are derived and analyzed in next sections.

### III. FREE ENERGY OF SL-MoS$_2$

First we consider the phases with $\eta_3 = 0$, corresponding to a SL-MoS$_2$, and potentially valid for SL-(MoS$_2$)$_{1-x}$/(ReS$_2$)$_x$ with small concentrations of Re, at x<<1. In this case, the free energy density (8) simplifies as

$$f_L[\mathbf{\eta}(\vec{r})] = \alpha_{11}\eta_\perp^2 + \beta_{11}\eta_\perp^4 + [\gamma_{111}\cos^2(3\varphi) + \gamma_{222}\sin^2(3\varphi)]\eta_\perp^6. \tag{9}$$

Minimization of Eq.(9) yields two solvable coupled equations for the polar angle $\varphi$ and absolute value $\eta_\perp$ of the in-plane components $\eta_1$ and $\eta_2$, which have the form:

$$\begin{cases}(\gamma_{111} - \gamma_{222})\sin(6\varphi) = 0, \\ \alpha_{11}\eta_\perp + 2\beta_{11}\eta_\perp^3 + 3[\gamma_{111}\cos^2(3\varphi) + \gamma_{222}\sin^2(3\varphi)]\eta_\perp^5 = 0.\end{cases} \tag{10}$$

The system Eq. (10) has 6 direction-degenerated minima ("states") corresponding to the vertices of regular hexagon the in the configuration space $\{\eta_1, \eta_2\}$, the parameters of which are listed in **Table SI** in **Appendix B**. The six global minima are separated by six local minima or saddles. All these 12 states we divided into "G$_1$-phase" with angles $\varphi_m = \frac{\pi}{3}m$ and rotated "G$_2$-phase" with $\varphi_m = \frac{\pi}{6} + \frac{\pi}{3}m$ ($m = 0, ..., 5$). The energies of the states in G$_1$- and G$_2$-phases differ only in the value of the nonlinearity parameter $\gamma_{iii}$. When $\gamma_{111} = \gamma_{222}$, the energies of G$_1$- and G$_2$-phases minima become indistinguishable and the equilibrium direction of $\eta_i$ in the $\{\eta_1,\eta_2\}$ plane in indeterminate. The six order parameters corresponding to the minima have the form $\eta_1 = \eta_\perp \cos(\varphi_m)$ and $\eta_2 = \eta_\perp \sin(\varphi_m)$ where $\varphi_m = \frac{\pi}{3}m$ for $G_1$-phase, and $\varphi_m = \frac{\pi}{6} + \frac{\pi}{3}m$ for $G_2$-phase, where $m = 0, ..., 5$. These hexagon-type solutions form two distinct groups corresponding to 2 triangles in the configuration space $\{\eta_1,\eta_2\}$. The first one is $\eta_1 = 0$, and $\eta_2 = \pm\eta_\perp$ (or $\eta_1 = \pm\eta_\perp$ and $\eta_2 = 0$), and the second one is $\eta_1 = \pm\frac{\eta_\perp}{2}$ and $\eta_2 = \pm\sqrt{3}\frac{\eta_\perp}{2}$ (or $\eta_2 = \pm\frac{\eta_\perp}{2}$ and $\eta_1 = \pm\sqrt{3}\frac{\eta_\perp}{2}$).

Similar to thermodynamic analysis of the perovskites, here we assume that only the coefficient $\alpha_{11}$ in Eq.(9) depends on Mo/Re atomic ratio $x$ in the following way

$$\alpha_{11}(x) = \alpha_{11}^0 a(x), \qquad a(x) = \frac{x}{x_{Mo}} - 1, \qquad 0 \le x << 1, \qquad 0 < x_{Mo}. \tag{11}$$

We further introduce the dimensionless variables, order parameters and expansion coefficients,

$$\tilde{g} = f_L \alpha_{11}^0 \sqrt{\frac{\alpha_{11}^0}{\gamma_{111}}}, \quad \tilde{\eta}_{1,2}^2 = \sqrt{\frac{\gamma_{111}}{\alpha_{11}^0}}\eta_{1,2}^2, \quad \eta_{1,2}^2 = \sqrt{\frac{\alpha_{11}^0}{\gamma_{111}}}\tilde{\eta}_{1,2}^2, \quad \beta = \beta_{11}\sqrt{\frac{\gamma_{111}}{\alpha_{11}^0}}, \quad \gamma = \frac{\gamma_{222}}{\gamma_{111}}, \tag{12}$$

As described above, $\alpha_{11}^0 > 0$, $\gamma_{111} \ge 0$ and $\gamma_{222} \ge 0$ to ensure thermodynamic stability. Using the designations Eq. (12) we derive the following expression for the dimensionless free energy density:

$$\tilde{g} = a(x)\tilde{\eta}_\perp^2 + \beta\tilde{\eta}_\perp^4 + \left[\cos^2(3\varphi) + \gamma\sin^2(3\varphi)\right]\tilde{\eta}_\perp^6. \tag{13}$$

The values of dimensionless order parameter components, corresponding energies (13) and their stability conditions are given in **Table I**.

**Table I.** Order parameter values in different phases of free energy (13), corresponding energies and their stability conditions

| Phase | Order parameter components $\tilde{\eta}_1$, $\tilde{\eta}_2$ | Phase energy and stability conditions |
|---|---|---|
| Disordered | $\tilde{\eta}_1 = 0$, $\tilde{\eta}_2 = 0$ | $\tilde{g} = 0 = \min$. Stability conditions are $a(x) \ge 0$, $\beta^2 - 4\gamma a(x) \le 0$ and $\beta^2 - 4a(x) \le 0$ |
| $G_2$-phase | $\tilde{\eta}_1 = \tilde{\eta}_\perp \cos(\varphi_m)$ $\tilde{\eta}_2 = \tilde{\eta}_\perp \sin(\varphi_m)$ $\tilde{\eta}_\perp = \sqrt{\frac{-a(x)}{\sqrt{\beta^2 - 3\gamma a(x)} + \beta}}$ $\varphi_m = \pm\frac{\pi}{6} + \frac{2\pi}{3}m$, $m = 0, 1, 2$. | $\tilde{g}_L^{G2} = \frac{-a^2(x)\left[\beta^2 - 4\gamma a(x)\right]}{2\left[\beta^2 - 3\gamma a(x)\right]^{3/2} + 2\beta^3 - 9\beta\gamma a(x)} = \min$ Stability conditions are $0 < \gamma < 1$ and $a(x) < 0$, or $a(x) > 0$, $0 < \gamma < 1$, $\beta^2 \ge 4a(x)\gamma$, $\beta < 0$. |

| $G_1$-phase | $\tilde{\eta}_1 = \tilde{\eta}_\perp \cos(\varphi_m)$  $\tilde{\eta}_2 = \tilde{\eta}_\perp \sin(\varphi_m)$ $\tilde{\eta}_\perp = \sqrt{\dfrac{-a(x)}{\sqrt{\beta^2 - 3a(x)} + \beta}}$ $\varphi_m = \dfrac{\pi}{3}m,\ m = 0, ..., 5.$ | $\tilde{g}_L^{G1} = \dfrac{-a^2(x)\left[\beta^2 - 4a(x)\right]}{2\left[\beta^2 - 3a(x)\right]^{3/2} + 2\beta^3 - 9\beta a(x)} = \min$ Stability conditions are $\gamma > 1$, $a(x) < 0$, or $\gamma > 1$, $a(x) > 0$, $\beta^2 \geq 4a(x)$, $\beta < 0$ |
|---|---|---|

Free energy of SL-(MoS$_2$) as a function of order parameter components $\tilde{\eta}_1$ and $\tilde{\eta}_2$ calculated for several values β, γ and *a*= −1 are shown in **Figs. 2**. Note the evolution of characteristic behaviors between the cases β= −1 and β= +1. **Figs. 2(a)-(c)** show the changes of the free energy relief with γ increase at negative β= −1. At γ<<1 free energy has 6 deep and narrow equivalent minima located in the ends of regular hexagon, separated by saddle points and a high central maxima ["deep" G$_2$-phase, **Fig. 2(a)**]. With γ increase the minima become shallow and wide, and eventually split into the ring in the limiting case γ=1 shown in **Fig. 2(b)**. For γ>1 the 6 minima gradually appear again and become deeper and narrower with further increase of γ. At γ>>1 free energy has 6 deep and thin equivalent minima located in the ends of regular hexagon, separated by saddle points and a high central maxima ["deep" G$_1$-phase, **Fig. 2(c)**]. Note that the minima position at γ>1 (G$_2$-phase) is rotated with respect to the case γ<1 (G$_1$-phase) at 30 degrees, as expected from the analytical results listed in **Table I**.

**Figs. 2(d)-(f)** show the changes of the free energy surface with γ increase at fixed positive β= −1. At γ<<1 free energy has 6 shallow and wide equivalent minima located in the ends of regular hexagon, connected by a ditch and separated with zero central maxima ["shallow" G$_2$-phase, **Fig. 2(d)**]. With γ increase the minima smear, and eventually split into the ring in the case γ=1, as shown in **Fig. 2(e)**. For γ>1 the 6 minima gradually appear again. At γ>>1 the free energy has 6 shallow and wide equivalent minima located in the ends of regular hexagon, rotated with respect to the case γ<1 at 30 degrees ["shallow" G$_1$-phase, **Fig. 2(f)**]. Note that the minima rotation is in agreement with analytical results listed in **Table I**.

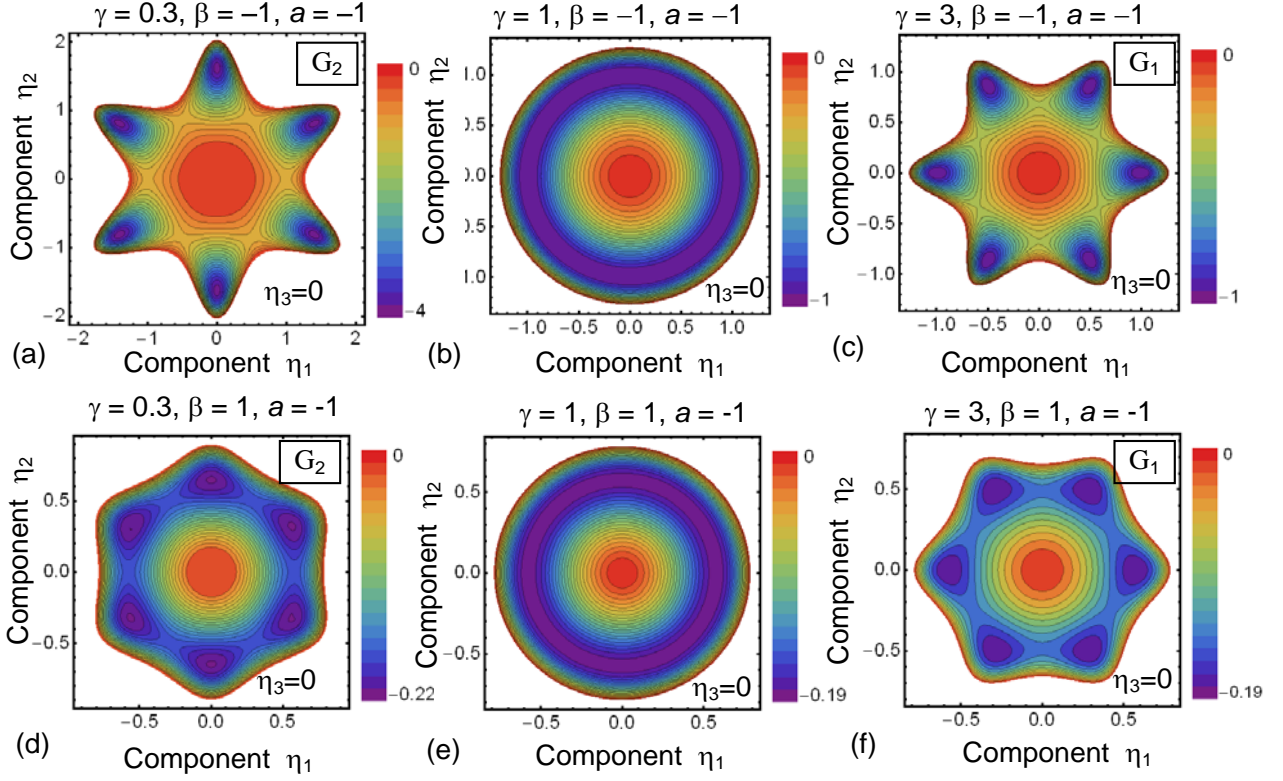

**FIGURE 2.** Free energy of SL-MoS$_2$ in dependence on order parameter components $\tilde{\eta}_1$ and $\tilde{\eta}_2$ calculated for several values of dimensionless parameters β, γ and *a* indicated above the plots **(a)-(f)**. The component $\tilde{\eta}_3 = 0$.

### IV. FREE ENERGY, PHASE DIAGRAMS AND ORDER PARAMETER OF SL-MoS$_2$-ReS$_2$

We further consider the phases with both $\eta_3 = 0$ and $\eta_3 \neq 0$ possible for a 2D SL-(MoS$_2$)$_{1-x}$-(ReS$_2$)$_x$, with nonzero fraction of Re atoms, i.e. at arbitrary $x > 0$. We assume that the inequalities $\gamma_{111} \geq 0$, $\gamma_{222} \geq 0$ and $\gamma_{333} \geq 0$ are valid, as required for the stability of the phase described by functional Eq. (5). In this case we performed numerical analysis using the same dimensionless variables as in the case $\eta_3 = 0$ [see Eqs.(12)], and additionally introduce dimensionless variables and parameters:

$$\tilde{g} = f_L \alpha_{11}^0 \sqrt{\frac{\alpha_{11}^0}{\gamma_{111}}}, \quad \tilde{\eta}_{1,2}^2 = \sqrt{\frac{\gamma_{111}}{\alpha_{11}^0}} \eta_{1,2}^2, \quad \tilde{\eta}_3^2 = \sqrt{\frac{\gamma_{111}}{\alpha_{11}^0}} \eta_3^2, \quad \gamma = \frac{\gamma_{222}}{\gamma_{111}}, \quad \lambda = \frac{\alpha_{33}^0}{\alpha_{11}^0}, \quad (14a)$$

$$\beta = \beta_{11} \sqrt{\frac{\gamma_{111}}{\alpha_{11}^0}}, \quad \tilde{\beta}_{13} = \beta_{13}^0 \sqrt{\frac{\gamma_{111}}{\alpha_{11}^0}}, \quad \tilde{\beta}_{15} = \beta_{15}^0 \sqrt{\frac{\gamma_{111}}{\alpha_{11}^0}}, \quad \delta = \beta_{33} \sqrt{\frac{\gamma_{111}}{\alpha_{11}^0}}. \quad (14b)$$

$$\tilde{\gamma}_{113} = \frac{\gamma_{113}}{\gamma_{111}}, \quad \tilde{\gamma}_{133} = \frac{\gamma_{133}}{\gamma_{111}}, \quad \tilde{\gamma}_{333} = \frac{\gamma_{333}}{\gamma_{111}}, \quad \tilde{\gamma}_{115}(x) = \frac{\gamma_{115}(x)}{\gamma_{111}}, \quad \tilde{\gamma}_{135}(x) = \frac{\gamma_{135}(x)}{\gamma_{111}}. \quad (14c)$$

The concentration dependent parameters in Eqs.(14a)-(14d) has the form,

$$\alpha_{11}(x) = \alpha_{11}^0 a(x), \qquad \alpha_{33}(x) = \alpha_{33}^0 c(x), \qquad \beta_{13}(x) = \tilde{\beta}_{13} b(x), \qquad \beta_{15}(x) = \tilde{\beta}_{15} b(x), \quad (14d)$$

$$a(x) = \frac{x}{x_{Mo}} - 1, \quad c(x) = 1 - \frac{x}{x_{Re}}, \quad b(x) = 1 - \exp\left(-\frac{x}{\Delta}\right) \approx \begin{cases} 0, & 0 < x \ll \Delta, \\ 1, & \Delta \ll x. \end{cases} \quad (14e)$$

where the characteristic concentrations $x_{Re}$, $x_{Mo}$ and dispersion $\Delta$ are positive. Hereinafter we regard that $\alpha_{11}^0 > 0$. The parameter $a(x)$ increases from "−1" for $x$=0 (MoS$_2$) to positive values for $x > x_{Mo}$ (closer to ReS$_2$). Parameter $c(x)$ decreases from "+1" for $x$=0 (MoS$_2$) to negative values for $x > x_{Re}$ (closer to ReS$_2$).

Parameter $b(x)$ describes the coupling strength between $\tilde{\eta}_1$, $\tilde{\eta}_2^2$ and $\tilde{\eta}_3^2$. The coupling is possible only when neighboring dopant (Re) atoms start to interact via e.g. changes of the host lattice force matrix (that is possible for $x > \Delta_{cr}$), and, later on with $x$ increase, possibly via percolation mechanism. To model the interaction, we regard that $b(x)$ should be very small is nearly zero for $0 < x \ll \Delta$, then it increases exponentially and saturates to unity for $x \gg \Delta$. Similarly to the coupling parameters $\beta_{15}(x)$ and $\beta_{13}(x)$, other $\tilde{\eta}_{1,2} - \tilde{\eta}_3$ coupling parameters $\tilde{\gamma}_{115}(x)$, $\tilde{\gamma}_{133}(x)$ and $\tilde{\gamma}_{135}(x)$ should be proportional to Re-fraction, $x$.

For the case $\delta > 0$ there is no need to consider the higher expansion terms proportional to $\tilde{\gamma}_{113}\eta_\perp^4\eta_3^2 + \tilde{\gamma}_{133}\eta_\perp^2\eta_3^4 + \tilde{\gamma}_{333}\eta_3^6$, since the functional stability with $\eta$ increase is provided by lower terms. So the simplification $\tilde{\gamma}_{113} = \tilde{\gamma}_{133} = \tilde{\gamma}_{333} = \tilde{\gamma}_{115} = \tilde{\gamma}_{135} = 0$ is grounded for $\delta > 0$ and regarded valid hereinafter for the sake of simplicity.

Taking into account all above assumptions and simplifications, the free energy density (8) in dimensionless variables (14) has the form:

$$\tilde{g} \approx \begin{pmatrix} a(x)\tilde{\eta}_\perp^2 + \beta\tilde{\eta}_\perp^4 + [\cos^2(3\varphi) + \gamma \sin^2(3\varphi)]\tilde{\eta}_\perp^6 + \\ \lambda c(x)\tilde{\eta}_3^2 + \delta\tilde{\eta}_3^4 + b(x)\tilde{\eta}_\perp^2 [\tilde{\beta}_{13}\tilde{\eta}_3^2 + \tilde{\beta}_{15}\tilde{\eta}_\perp \tilde{\eta}_3 \cos(3\varphi)] \end{pmatrix} \quad (15)$$

Minimization of Eq.(15) yields three coupled equations for $\tilde{\eta}_i$ and $\varphi$:

$$\begin{cases} [2(\gamma-1)\tilde{\eta}_\perp^6 \cos(3\varphi) - \tilde{\beta}_{15} b(x)\tilde{\eta}_\perp^3 \tilde{\eta}_3]\sin(3\varphi) = 0, \\ 2a(x)\tilde{\eta}_\perp + 4\beta\tilde{\eta}_\perp^3 + 6[\cos^2(3\varphi) + \gamma\sin^2(3\varphi)]\tilde{\eta}_\perp^5 + b(x)[2\tilde{\beta}_{13}\tilde{\eta}_3^2\tilde{\eta}_\perp + 3\tilde{\beta}_{15}\tilde{\eta}_\perp^2\tilde{\eta}_3 \cos(3\varphi)] = 0, \\ 2(\lambda c(x) + \tilde{\beta}_{13} b(x)\tilde{\eta}_\perp^2)\tilde{\eta}_3 + 4\delta\tilde{\eta}_3^3 + \tilde{\beta}_{15} b(x)\tilde{\eta}_\perp^3 \cos(3\varphi) = 0. \end{cases} \quad (16)$$

The polar angles of global and local minima (or saddles) can be found from equations, $\sin(3\varphi) = 0$ and $\cos(3\varphi) = \dfrac{\tilde{\beta}_{15} b \tilde{\eta}_3}{2(\gamma-1)\tilde{\eta}_\perp^3}$, each of which has 6 roots at the polar circle. Approximate solutions of Eqs.(16) are derived in **Appendix B**.

For several phases, Equations (16) can be solved analytically, and corresponding expressions for $\tilde{\eta}_i$ and $\tilde{g}(\tilde{\eta}_i)$ are listed in **Table II**. It is seen from the table that disordered $G_0$, and long-range ordered $G_2$, $G_1$, $G_3$, $G_{23}$, $G_{13}$, and $G_{123}$ phases can exist for a certain ranges of material parameters. The properties of these phases are described below.

$G_2$ and $G_1$ phases with $\tilde{\eta}_3 = 0$ correspond to 6-fold energy degenerated states. Hence, 6 types of structural in-plane domains exist. 2-fold degenerated $G_3$ phase with $\tilde{\eta}_1 = \tilde{\eta}_2 = 0$ contains 2 types of structural out-of-plane domains with $\tilde{\eta}_3 = \pm\sqrt{-\dfrac{\lambda c(x)}{2\delta}}$. 12-fold degenerate $G_{23}$ phases contain 12 types of structural domains (direct product of 6 in-plane and 2 out-of-plane ones) for $\beta_{15} = 0$. Analytical expressions corresponding to $G_{13}$ phase [listed in **Table II**] are exact only for $\beta_{15} = 0$, when it has 12-fold degenerated energy minima located it 2 regular hexagons located in plane $\{\tilde{\eta}_1, \tilde{\eta}_2\}$ with $\tilde{\eta}_3 = \pm\sqrt{-\dfrac{\lambda c + \tilde{\beta}_{13} b \tilde{\eta}_\perp^2}{2\delta}}$.

For $\beta_{15} = 0$, the inversion with respect to the sign of $\tilde{\eta}_i$, $\tilde{g}(+\tilde{\eta}_1,\tilde{\eta}_2,\tilde{\eta}_3) \equiv \tilde{g}(-\tilde{\eta}_1,\tilde{\eta}_2,\tilde{\eta}_3)$, $\tilde{g}(\tilde{\eta}_1,-\tilde{\eta}_2,\tilde{\eta}_3) \equiv \tilde{g}(\tilde{\eta}_1,+\tilde{\eta}_2,\tilde{\eta}_3)$ and $\tilde{g}(\tilde{\eta}_1,\tilde{\eta}_2,+\tilde{\eta}_3) \equiv \tilde{g}(\tilde{\eta}_1,\tilde{\eta}_2,-\tilde{\eta}_3)$, exists in all above phases. If $\tilde{\beta}_{15} \neq 0$, the degeneracy for $\tilde{\eta}_1$ is lifted, $\tilde{g}(+\tilde{\eta}_1,\tilde{\eta}_2,\tilde{\eta}_3) \neq \tilde{g}(-\tilde{\eta}_1,\tilde{\eta}_2,\tilde{\eta}_3)$. However, the equality $\tilde{g}(+\tilde{\eta}_1,\tilde{\eta}_2,\tilde{\eta}_3) = \tilde{g}(-\tilde{\eta}_1,\tilde{\eta}_2,-\tilde{\eta}_3)$ is valid for $\tilde{\beta}_{15} \neq 0$, indicating that the change of $\tilde{\eta}_3$ sign is accompanied with its 6 minima rotation at 60-degree in the plane $\{\tilde{\eta}_1, \tilde{\eta}_2\}$.

Importantly, the term $\tilde{\beta}_{15} b(x) \tilde{\eta}_1 \tilde{\eta}_3 (\tilde{\eta}_1^2 - 3\tilde{\eta}_2^2)$ with nonzero amplitude $\tilde{\beta}_{15} b(x)$ can induce the stable "mixed" phases. In particular, $G_{23}$ phase continuously transforms in the 12-fold degenerated "mixed" $G_{123}$ at $\tilde{\beta}_{15} \neq 0$, and corresponding exact expressions for the order parameters and energy are also listed in **Table II**. Positive value of $\tilde{\eta}_3 = +\sqrt{-\dfrac{\lambda c + \tilde{\beta}_{13} b \tilde{\eta}_\perp^2}{2\delta} - \dfrac{\tilde{\beta}_{15}^2 b^2}{4\delta(\gamma-1)}}$ corresponds to 6 stable minima (separated by 3 metastable states and 3 saddles), and negative $\tilde{\eta}_3 = -\sqrt{-\dfrac{\lambda c + \tilde{\beta}_{13} b \tilde{\eta}_\perp^2}{2\delta} - \dfrac{\tilde{\beta}_{15}^2 b^2}{4\delta(\gamma-1)}}$ corresponds 60-degree rotated 6 stable minima of the same depth, since $\tilde{g}(+\tilde{\eta}_1,\tilde{\eta}_2,+\tilde{\eta}_3) = \tilde{g}(-\tilde{\eta}_1,\tilde{\eta}_2,-\tilde{\eta}_3)$. The minima can be shallow depending on the absolute value of coupling parameter $|\tilde{\beta}_{15} b(x)|$.

For small coupling strength $0 < |\tilde{\beta}_{15} b| \ll 1$, $0 < |\tilde{\beta}_{13} b| \ll 1$ the properties of $G_{13}$ phase can be analyzed within perturbation theory. In this case, the minima corresponding to the phase acquire

properties qualitatively similar to the ones inherent to the "mixed" $G_{123}$ phase. In particular, for $\tilde{\beta}_{15} \neq 0$ $G_{13}$ phase can be either "proper", if $\lambda c < 0$, or "incipient" if $\lambda c > 0$. The "proper" phase $G_{13}$ is 6-fold degenerated for the each sign of $\tilde{\eta}_3$ (positive or negative), indicating the transition of a planar hexagonal-type 6 minima to two triangular-type 3 global minima located in different planes $\tilde{\eta}_3 \approx +\sqrt{-\frac{\lambda c}{2\delta}}$ for $(-1)^m \tilde{\beta}_{15} b < 0$ and $\tilde{\eta}_3 \approx -\sqrt{-\frac{\lambda c}{2\delta}}$ for $(-1)^m \tilde{\beta}_{15} b > 0$. At arbitrary $\tilde{\beta}_{15} \neq 0$ $G_{13}$ phase has 3-fold energy degenerated in-plane deep minima separated either by 3 shallower minima, or 3 saddles. Thus 6 stable and 6 metastable domain configurations, required to recover the initial symmetry in accordance with Curie principle, should exist in $G_{13}$ phase.

**Table II.** Order parameter values in different phases of free energy (15b), corresponding energies and their stability conditions at $\delta > 0$ and $\tilde{\gamma}_{113} = \tilde{\gamma}_{133} = \tilde{\gamma}_{333} = \tilde{\gamma}_{115} = \tilde{\gamma}_{135} = 0$

| Phase | Order parameter components $\tilde{\eta}_1$, $\tilde{\eta}_2$ | Phase energy and stability conditions |
|---|---|---|
| Disordered $G_0$ | $\tilde{\eta}_1 = 0$, $\tilde{\eta}_2 = 0$, $\tilde{\eta}_3 = 0$ | $\tilde{g} = 0 = \min$. Stability conditions are $a(x) \geq 0$, $\beta^2 - 4\gamma a(x) \leq 0$ and $\beta^2 - 4a(x) \leq 0$, $\lambda c(x) > 0$ |
| In-plane $G_2$-phase | $\tilde{\eta}_1 = \tilde{\eta}_\perp \cos(\varphi_m)$, $\tilde{\eta}_2 = \tilde{\eta}_\perp \sin(\varphi_m)$, $\tilde{\eta}_\perp = \sqrt{\dfrac{-a(x)}{\sqrt{\beta^2 - 3\gamma a(x)} + \beta}}$ $\varphi_m = \pm\dfrac{\pi}{6} + \dfrac{2\pi}{3}m$, $m = 0, 1, 2$, $\tilde{\eta}_3 = 0$ | $\tilde{g}_L^{G2} = \dfrac{-a^2(x)[\beta^2 - 4\gamma a(x)]}{2[\beta^2 - 3\gamma a(x)]^{3/2} + 2\beta^3 - 9\beta\gamma a(x)} = \min$ Stability conditions are $0 < \gamma < 1$, $\lambda c(x) > 0$, $a(x) < 0$, or $0 < \gamma < 1$, $\lambda c(x) > 0$, $a(x) > 0$, $\beta^2 \geq 4a(x)\gamma$, $\beta < 0$. |
| In-plane $G_1$-phase | $\tilde{\eta}_1 = \tilde{\eta}_\perp \cos(\varphi_m)$, $\tilde{\eta}_2 = \tilde{\eta}_\perp \sin(\varphi_m)$, $\tilde{\eta}_\perp = \sqrt{\dfrac{-a(x)}{\sqrt{\beta^2 - 3a(x)} + \beta}}$ $\varphi_m = \dfrac{\pi}{3}m$, $m = 0, ..., 5$, $\tilde{\eta}_3 = 0$ | $\tilde{g}_L^{G1} = \dfrac{-a^2(x)[\beta^2 - 4a(x)]}{2[\beta^2 - 3a(x)]^{3/2} + 2\beta^3 - 9\beta a(x)} = \min$ Stability conditions are $\gamma > 1$, $\lambda c(x) > 0$, $a(x) < 0$, or $\gamma > 1$, $\lambda c(x) > 0$, $a(x) > 0$, $\beta^2 \geq 4a(x)$, $\beta < 0$ |
| Out-of-plane $G_3$-phase | $\tilde{\eta}_1 = \tilde{\eta}_2 = 0$, $\tilde{\eta}_3 = \pm\sqrt{-\dfrac{\lambda c(x)}{2\delta}}$ | $\tilde{g}_L^{G3} = -\dfrac{\lambda c(x)}{2\delta} = \min$, Stability conditions are $\lambda c(x) < 0$, $a(x) > 0$, $\beta > 0$ |
| $G_{23}$-phase Becomes $G_{123}$-phase at $\beta_{15} \neq 0$ | $\tilde{\eta}_1 = \tilde{\eta}_\perp \cos(\varphi_m)$, $\tilde{\eta}_2 = \tilde{\eta}_\perp \sin(\varphi_m)$, $\tilde{\eta}_\perp = \sqrt{\dfrac{1}{3\gamma}\left(\sqrt{\beta^{*2} - 3\gamma a^*} - \beta^*\right)}$ $\varphi_m = \pm\dfrac{1}{3}\arccos\left[\dfrac{\tilde{\beta}_{15} b \tilde{\eta}_3}{2(\gamma - 1)\tilde{\eta}_\perp^3}\right] + \dfrac{2\pi m}{3}$, $m = 0, 1, 2$ | $\tilde{g}_L^{G23} = \dfrac{-a^{*2}(\beta^{*2} - 4\gamma a^*)}{2(\beta^{*2} - 3\gamma a^*)^{3/2} + 2\beta^{*3} - 9\gamma\beta^* a^*} = \min$ $a^* = a - \lambda c\dfrac{\tilde{\beta}_{13} b}{2\delta} - \dfrac{\tilde{\beta}_{13}\tilde{\beta}_{15}^2 b^3}{4\delta(\gamma - 1)}$ and $\beta^* = \beta - \dfrac{\tilde{\beta}_{13}^2 b^2}{4\delta}$ Stability conditions are $0 < \gamma < 1$, $a^* < 0$, or |
| | $\tilde{\eta}_3 = \pm\sqrt{-\dfrac{\lambda c + \tilde{\beta}_{13}b\tilde{\eta}_\perp^2}{2\delta} - \dfrac{\tilde{\beta}_{15}^2 b^2}{4\delta(\gamma-1)}}$. | $0 < \gamma < 1$, $\lambda c < 0$, $a^* > 0$, $\beta^2 \geq 4a^*\gamma$, $\beta^* < 0$. |
| **G$_{13}$-phase** Exact expressions for $\beta_{15} = 0$ | $\tilde{\eta}_1 = \tilde{\eta}_\perp \cos(\varphi_m)$, $\tilde{\eta}_2 = \tilde{\eta}_\perp \sin(\varphi_m)$ $\varphi_m = \dfrac{\pi}{3}m$, $m = 0, ..., 5$ $\tilde{\eta}_\perp = \sqrt{\dfrac{1}{3}\left(\sqrt{\beta^{*2} - 3a^*} - \beta^*\right)}$ $\tilde{\eta}_3 = \pm\sqrt{-\dfrac{\lambda c + \tilde{\beta}_{13}b\tilde{\eta}_\perp^2}{2\delta}}$ | $\tilde{g}_L^{G13} = \dfrac{-a^{*2}\left(\beta^{*2} - 4a^*\right)}{2\left(\beta^{*2} - 3a^*\right)^{3/2} + 2\beta^{*3} - 9\beta^*a^*} = \min$ $a^* = a - \lambda c\dfrac{\tilde{\beta}_{13}b}{2\delta}$ and $\beta^* = \beta - \dfrac{\tilde{\beta}_{13}^2 b^2}{4\delta}$ Stability conditions: $\gamma > 1$, $\lambda c < 0$, $a^* < 0$, or $0 < \gamma < 1$, $a^* > 0$, $\beta^2 \geq 4a^*$, $\beta^* < 0$. |
| "proper" **G$_{13}$-phase** Approximate expressions for $\|\tilde{\beta}_{15}b\| << 1$, $\|\tilde{\beta}_{13}b\| << 1$ and $\lambda c < 0$ | $\tilde{\eta}_1 = \tilde{\eta}_\perp \cos(\varphi_m)$, $\tilde{\eta}_2 = \tilde{\eta}_\perp \sin(\varphi_m)$ $\varphi_m = \dfrac{\pi}{3}m$, $m = 0, ..., 5$ $\tilde{\eta}_\perp \approx \sqrt{\dfrac{1}{3}\left(\sqrt{\beta^{*2} - 3a^*} - \beta^*\right)}$ $\tilde{\eta}_3 \approx \pm\sqrt{-\dfrac{\lambda c + \tilde{\beta}_{13}b\tilde{\eta}_\perp^2}{2\delta} + \dfrac{(-1)^m \tilde{\beta}_{15}b\tilde{\eta}_\perp^3}{4\lambda c}}$ where $a^* = a - \lambda c\dfrac{\tilde{\beta}_{13}b}{2\delta}$, $\beta^* = \beta - \dfrac{\tilde{\beta}_{13}^2 b^2}{4\delta}$ and $\lambda c < 0$ | $\tilde{g}_L^{G13} \approx \dfrac{-a^{*2}\left(\beta^{*2} - 4a^*\right)}{2\left(\beta^{*2} - 3a^*\right)^{3/2} + 2\beta^{*3} - 9\beta^*a^*} +$ $\pm (-1)^m \dfrac{\tilde{\beta}_{15}b}{3\sqrt{3}}\left(\sqrt{\beta^{*2} - 3a^*} - \beta^*\right)^{3/2}\sqrt{-\dfrac{\lambda c}{2\delta}}$ is 6-fold degenerated for the each sign of $\tilde{\eta}_3$ (positive or negative), indicating the transition of a planar hexagon minima to 2 triangles minima located in different planes $\tilde{\eta}_3 \approx +\sqrt{-\dfrac{\lambda c}{2\delta}}$ for $(-1)^m \tilde{\beta}_{15}b < 0$ and $\tilde{\eta}_3 \approx -\sqrt{-\dfrac{\lambda c}{2\delta}}$ for $(-1)^m \tilde{\beta}_{15}b > 0$.. |
| "incipient" **G$_{13}$-phase** Approximate expressions for $\|\tilde{\beta}_{15}b\| << 1$, $\|\tilde{\beta}_{13}b\| << 1$ | $\tilde{\eta}_1 = \tilde{\eta}_\perp \cos(\varphi_m)$, $\tilde{\eta}_2 = \tilde{\eta}_\perp \sin(\varphi_m)$ $\varphi_m = \dfrac{\pi}{3}m$, $m = 0, ..., 5$ $\tilde{\eta}_\perp \approx \sqrt{\dfrac{\sqrt{\beta^2 - 3a(1-\eta)} - \beta}{3(1-\eta)}}$ $\tilde{\eta}_3 \approx -\dfrac{(-1)^m \tilde{\beta}_{15}b\tilde{\eta}_\perp^3}{2\lambda c}$ | $\tilde{g}_L^{G13} = \dfrac{-a^2\left(\beta^2 - 4(1-\eta)a\right)}{2\left(\beta^2 - 3(1-\eta)a\right)^{3/2} + 2\beta^3 - 9(1-\eta)\beta a}$ $-\dfrac{\tilde{\beta}_{15}^2 b^2 \tilde{\eta}_\perp^4}{2\left(\lambda c + \tilde{\beta}_{13}b\tilde{\eta}_\perp^2\right)}$ where $\eta = \dfrac{\tilde{\beta}_{15}^2 b^2}{4\lambda c}$ and $\lambda c > 0$ |

Below we will analyze the phase diagrams and order parameter values calculated numerically for arbitrary $\tilde{\beta}_{15}$ and $\tilde{\beta}_{13}$ values. The continuous transition from 6 equivalent potential wells (corresponding to the case $b(x) = 0$) to the 3 equivalent deep potential wells separated by either 3 shallow wells, or by 3 saddles, or by 3 local maxima, occurs with $b(x)$ increase from "0" to "1", fixed other parameters and positive $\tilde{\eta}_3$ (see **Figs. 3-4)**. For negative $\tilde{\eta}_3$ there are other 3 equivalent deep potential wells separated by either 3 shallow wells, or by 3 saddles, or by 3 local maxima. The transition corresponds to point symmetry changes from in-plane hexagon to the spatially separated 2 rotated triangles induced by Re doping. Hence, the transition reflects the situation when one "moves" along x-axes from SL-MoS$_2$ with $x \approx 0$ to the morphotropic boundary at $x \geq 0.5$, and then approaches SL-ReS$_2$ at $x = 1$.

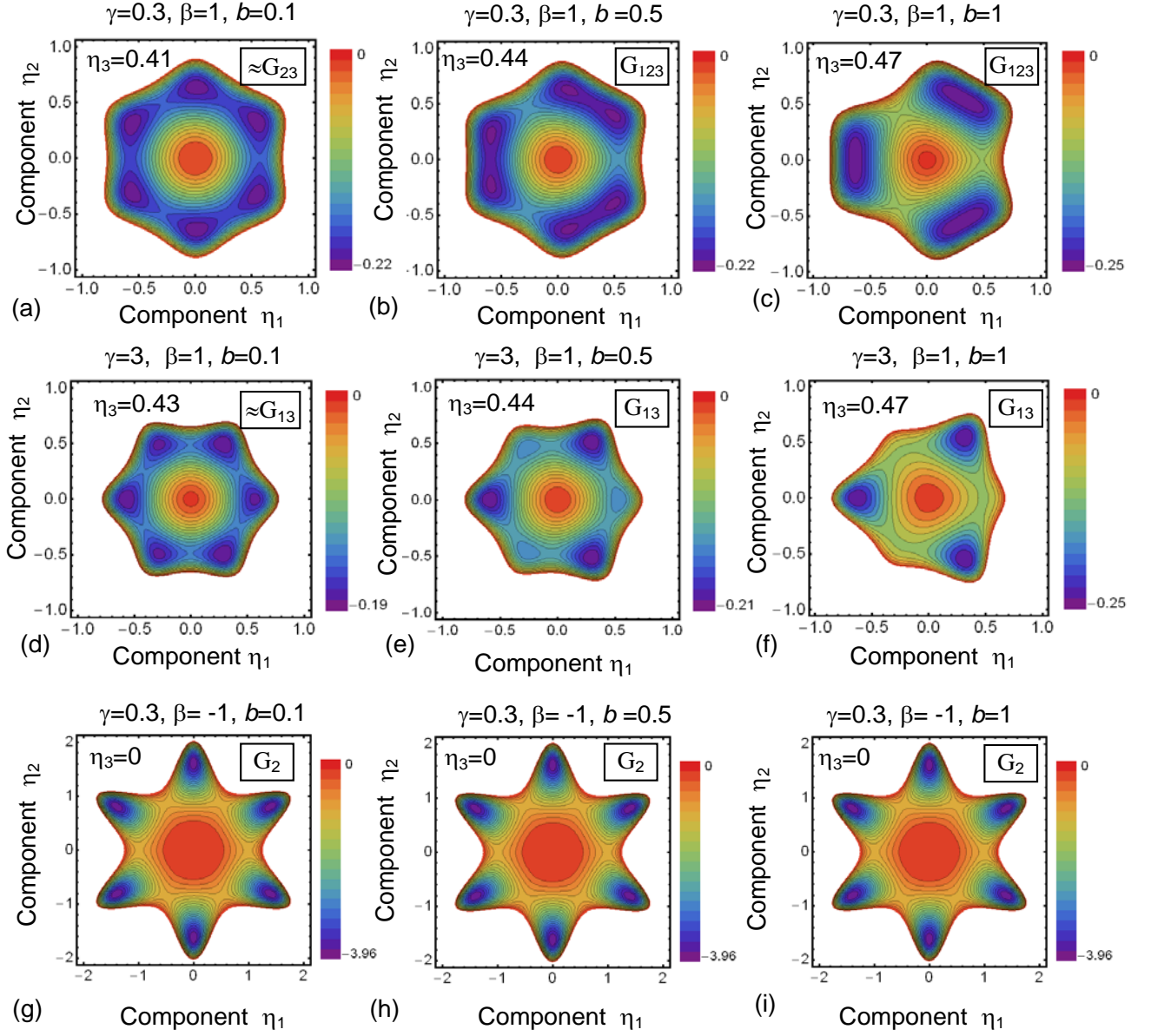

**FIGURE 3.** Free energy of SL-(MoS$_2$)$_{1-x}$-(ReS$_2$)$_x$ in dependence on order parameter components $\tilde{\eta}_1$ and $\tilde{\eta}_2$ calculated for several values of dimensionless parameters $\beta$, $\gamma$, and $b$ indicated above the plots **(a)-(i)**. The positive (or zero) $\tilde{\eta}_3$ values, which correspond to the energy minima, are indicated inside the plots. The in-plane rotated patterns corresponding to negative $\tilde{\eta}_3$ values are not shown. Parameters $a = -0.95$, $\tilde{\beta}_{13} = 1$, $\tilde{\beta}_{15} = 1$, $\lambda c = -1$ and $\delta = 2$.

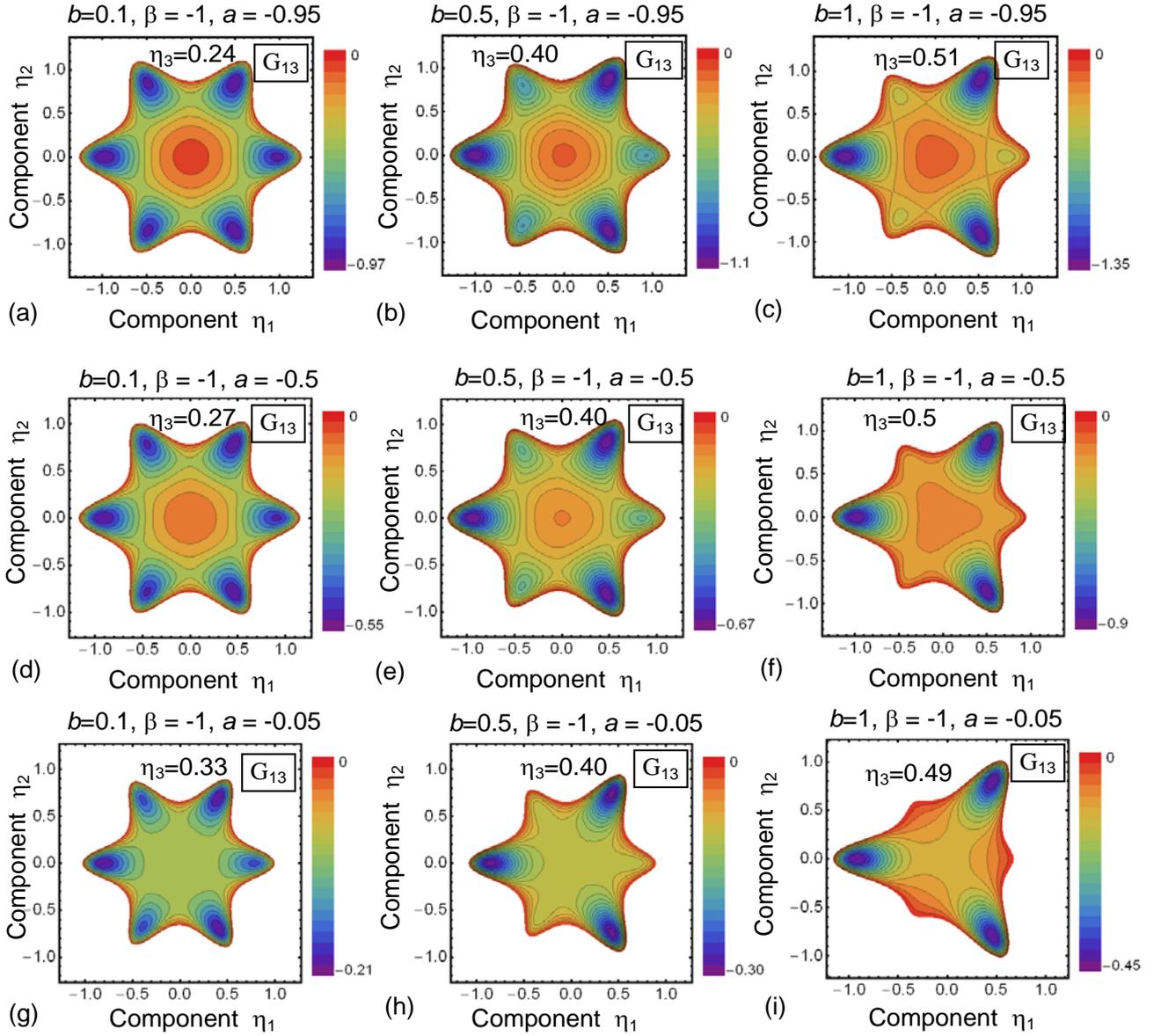

**FIGURE 4.** Free energy of SL-$(MoS_2)_{1-x}$-$(ReS_2)_x$ in dependence on order parameter components $\tilde{\eta}_1$ and $\tilde{\eta}_2$ calculated for several values of dimensionless parameters $\beta$, $b$ and $a$ indicated above the plots **(a)-(i)**. The positive (or zero) $\tilde{\eta}_3$ values, which correspond to the energy minima, are indicated inside the plots. The in-plane rotated patterns corresponding to negative $\tilde{\eta}_3$ values are not shown. Parameters $\gamma=3$ and $\beta=-1$; others are the same as in **Fig. 3.**

The particular form of this **"6"-to-"2×3"** or **"hexagon"-to-"two rotated triangles"** transition depends on the signs and values of other parameters, such as $\gamma$ [compare top **(a)-(c)** and bottom **(d)-(f)** plots in **Fig. 3**, calculated for $\gamma = 0.3$ and $\gamma = 3$, respectively], $\beta$ [compare plots in **Fig. 3** for $\beta=1$ with the plots in **Fig. 4** for $\beta = -1$] and $a$ [compare top **(a)-(c),** middle **(d)-(f)** and bottom **(g)-(i)** plots in **Figs. 4**, calculated for $a = -0.95, -0.5$ and $-0.05$, respectively].

Note that hexagon-shaped free energy surfaces in **Figs. 3(g)-(i)** calculated for negative β = −1, $a$ = −0.95 and γ = 0.3 are almost insensitive to increase of $b$. Namely, both position and depth of 6 equivalent minima remain unchanged when $b$ increases from "0" to "1", since $\tilde{\eta}_3 = 0$ for the case. The unchanged hexagons shown in **Figs. 3(g)-(i)** contrast with the hexagon-to-triangles transition shown in plots **Figs. 3(a)-(c)**, where $\tilde{\eta}_3 \neq 0$. The pronounced hexagon-to-triangles transition [evidently seen in **Figs. 4(a)-(c), (d)-(f),** and **(g)-(i)** along horizontal lines] occur with $b$ increase, and are accompanied by $\tilde{\eta}_3$ increase from 0.2 to 0.5. Hence we conclude that the appearance of nonzero $\tilde{\eta}_3$ induces the hexagon-to-triangles transition.

It should be noted that all energy maps shown in **Figs. 3-4** for $\tilde{\eta}_3 = 0$ or $\tilde{\eta}_3 > 0$, are characterized by the same (or very close) angular positions of global (or local) minima, which amount varies from 6 to 3 depending on $b$ value and β sign. In other words, the minima and saddles positions appeared weakly $b$-independent, while the minima depth and corresponding values of the order parameter components can significantly depend on $b$ (exception is the case β = −1 and γ <1). We further established that the hexagon-to-triangles transition is continuous, i.e. when it happens, it happens gradually with $b$ increase.

These two circumstances, namely the weak dependence of minima angular position on $b$ and the hexagon-to-triangles transition continuity, allows us to conclude that the increase of the term $\tilde{\beta}_{15}b(x)\tilde{\eta}_1\tilde{\eta}_3(\tilde{\eta}_1^2 - 3\tilde{\eta}_2^2)$ in the free energy Eq.(7) is responsible for the for the "hexagon"-to-"two rotated triangles" transition. Very small $b$ influences the appearance and stability of G$_1$- and G$_2$-phases only weakly. Hence, the term can be considered as "perturbation", and the fact allows us to check the numerical algorithm for the phase diagrams and order parameter calculation using **Table III** for $\tilde{\beta}_{15} = 0$, and $0 < |\tilde{\beta}_{15}b| \ll 1$.

## A. The case $\tilde{\beta}_{15} = 0$

First we analyze phase diagrams for $\tilde{\beta}_{15} = 0$ as shown in **Figs. 5(a)-(c)**. Color images **Figs. 5(a)** schematically show the distribution of the free energy minima in the plane $\{\tilde{\eta}_1, \tilde{\eta}_2\}$ at fixed $\tilde{\eta}_3$ corresponding to the minima of corresponding phase. The profiles of the order parameters corresponding to the cut-lines of the diagrams at fixed γ and $\tilde{\beta}_{13}$ are shown in **Figs. 6(a)-(c)**, respectively. The amplitude $\tilde{\eta}_\perp$ of six in-plane roots $\tilde{\eta}_1 = \tilde{\eta}_\perp \cos(\varphi_m)$ and $\tilde{\eta}_2 = \tilde{\eta}_\perp \sin(\varphi_m)$, and the positive root for $\tilde{\eta}_3$ are shown in the figures.

The phase diagram of SL-$(MoS_2)_{1-x}$-$(ReS_2)_x$ dependent on Re/Mo fraction $x$ and nonlinearity coefficient $\gamma$ are shown in **Fig. 5(a).** $G_2$-phase is stable at $\gamma<1$ and $x<x_C$ ($x_C \approx 0.51$). $G_1$-phase is stable at $x<x_C$ and $\gamma>1$. $G_{23}$-phase is stable at $\gamma<1$ and $x_C<x<x_M$ ($x_M \approx 0.91$). $G_{13}$-phase is stable at $x_C<x<x_M$ and $\gamma>1$. $G_3$-phase is stable at $x>x_M$. Note that the value of $x_C \approx 0.51$ is very close to $x_{Re}=0.5$, and the value $x_M \approx 0.91$ is close to $x_{Mo}=1$. The difference between $x_C$ and $x_{Re}$ is related with nonzero $\tilde{\beta}_{13}=0.1$, and it increases with $\tilde{\beta}_{13}$ increase. The phase boundaries in **Fig.5(a)** are either almost vertical ($G_2$-$G_{23}$, $G_1$-$G_{13}$, $G_{23}$-$G_3$ and $G_{13}$-$G_3$ boundaries) or horizontal ($G_2$-$G_1$ and $G_{23}$-$G_{13}$ boundaries).

The dependence of the order parameter components $\tilde{\eta}_\perp$ and $\tilde{\eta}_3$ on $x$, calculated for $\tilde{\beta}_{13}=0.1$, $\gamma=3$ (solid curves) or $\gamma=0.3$ (dashed curves) are shown in **Fig.6(a).** The dependences demonstrate the second order phase transition between $G_1$ and $G_{13}$ (or $G_2$ and $G_{23}$) phases at $x=x_C$, when the out-plane component $\tilde{\eta}_3$ appears and is almost the same for $\gamma=3$ and $\gamma=0.3$ (black dashed and solid curves coincide). Then another second order phase transition between $G_{13}$ and $G_3$ ($G_{23}$ and $G_3$) phases happens at $x=x_M$, when in-plane components $\tilde{\eta}_{1,2}$ (red solid curve for $\gamma>1$ or blue dashed curve for $\gamma<1$) disappear.

The phase diagrams of SL-$(MoS_2)_{1-x}$-$(ReS_2)_x$, in dependence on Re/Mo fraction $x$ and nonlinearity coefficient $\tilde{\beta}_{13}$ is shown in **Figs. 5(b)** and **5(c)**, calculated for $\gamma=3$ and $\gamma=0.3$, respectively. $G_1$, $G_{13}$, and $G_3$ are stable for $\gamma=3$ [**Fig. 5(b)**]. $G_2$, $G_{23}$, and $G_3$ are stable for $\gamma=0.3$ [**Fig. 5(c)**]. **Fig. 5(c)** looks almost the same as **Fig. 5(b)** with substitution of the subscript "1" for $\gamma>1$ to subscript "2" for $\gamma<1$. The result is expected because $\gamma$ values for the figures are close to inverse ones, $1/3\approx0.3$. The phase boundaries between all three phases in these figures are curved, and for chosen parameters three phases meet in a tricritical point with coordinates $x \approx 0.62$ and $\tilde{\beta}_{13} \approx 1.6$. $G_1$ (or $G_2$) phase is stable at $x<0.4$, and the region of its stability increases up to $x<0.62$ with $\tilde{\beta}_{13}$ increase from $-1$ to 2. $G_{13}$ (or $G_{23}$) phase is stable at $x>0.4$, and the region of its stability decreases with $\tilde{\beta}_{13}$ increase up to disappearance in the tricritical point $\tilde{\beta}_{13} \approx 1.6$. $G_3$ phase is stable at $x>0.62$ and $\tilde{\beta}_{13}>0$, and the region of its stability increases with $\tilde{\beta}_{13}$ increase.

The dependence of the order parameter components on $x$, calculated at $\gamma=3$, and relatively high positive $\tilde{\beta}_{13}=+1$ (solid curves) or negative $\tilde{\beta}_{13}=-1$ (dashed curves) are shown in **Fig. 6(b)**. The dependences demonstrate the second order phase transition between $G_1$ and $G_{13}$ phases at $x \approx 0.4$ (for $\tilde{\beta}_{13} = -1$) or at $x \approx 0.59$ (for $\tilde{\beta}_{13} = +1$), when the out-plane component $\tilde{\eta}_3$ appears and rapidly increases with $x$ (compare black dashed and solid curves). Another second order phase transition between G$_{13}$ and G$_3$ phases happens at $x \approx 0.68$ and $\tilde{\beta}_{13} = +1$, when in-plane components with amplitude $\tilde{\eta}_\perp$ (red solid curve) disappears. The transition between G$_{13}$ and G$_3$ phases is absent for $\tilde{\beta}_{13} = -1$ (red dashed curve for $\tilde{\eta}_\perp$ exists for all $x$). The features (kinks) on the $\tilde{\eta}_\perp$, and $\tilde{\eta}_3$ curves correspond to the phase transition points (appearance of $\tilde{\eta}_3$ and $\tilde{\eta}_\perp$, respectively). The situation shown in **Fig. 6(b)** is typical for relatively high $|\tilde{\beta}_{13}|$.

The dependence of the order parameter components on $x$, calculated for $\gamma = 0.3$, and relatively small positive $\tilde{\beta}_{13} = +0.1$ (solid curves) or negative $\tilde{\beta}_{13} = -0.1$ (dashed curves) are shown in **Fig. 6(c)**. The dependences demonstrate the second order phase transition between G$_2$ and G$_{23}$ phases at $x \approx x_C$, when the out-plane component $\tilde{\eta}_3$ appears and looks very similar for $\tilde{\beta}_{13} = +0.1$ and $\tilde{\beta}_{13} = -0.1$ (black dashed and solid curves are very close). Another second order phase transition between G$_{23}$ and G$_3$ phases happens at $x \approx x_M$ and $\tilde{\beta}_{13} = +0.1$, when in-plane component $\tilde{\eta}_\perp$ (blue solid curve) disappears. The transition between G$_{23}$ and G$_3$ phases is absent for $\tilde{\beta}_{13} = -0.1$ (blue dashed curve for $\tilde{\eta}_2$ exists for all $x$).

## B. The case $\tilde{\beta}_{15} \neq 0$

We further explore phase diagrams of SL-(MoS$_2$)$_{1-x}$-(ReS$_2$)$_x$ for $\tilde{\beta}_{15} \neq 0$ as shown in **Figs. 5(d)-(f)**. Small color images **Figs. 5(d)** and **5(e)** schematically show the distribution of the free energy minima in the plane $\{\tilde{\eta}_1, \tilde{\eta}_2\}$ at fixed $\tilde{\eta}_3$ corresponding to the stability of corresponding phase. The profiles of the order parameters corresponding to the cut-lines of the diagrams at fixed $\tilde{\beta}_{15}$ are shown in **Figs. 6(d)** and **7(a-d)**, respectively.

The phase diagrams of SL-(MoS$_2$)$_{1-x}$-(ReS$_2$)$_x$ as a function of concentration $x$ and nonlinearity coefficient $\tilde{\beta}_{15}$, for different $\gamma$ and $\tilde{\beta}_{13}$ are shown in **Fig. 5(d)-(f)**. For $\gamma = 3$ and $\tilde{\beta}_{13} = 0.1$, a very thin region of G$_1$ phase has a wedge-like shape; it is surrounded by G$_{13}$ phase, that is followed by G$_3$ phase at $x > x_M$ and [**Fig. 5(d)**]. For $\gamma = 0.3$ and $\tilde{\beta}_{13} = 0.1$, a wide region of G$_2$ phase exists at $x < x_C$; it is followed by a beak-shape region of G$_{123}$ phase, surrounded by G$_{13}$ phase, that is followed by a segment of G$_3$ phase at $x > x_M$ and $\tilde{\beta}_{15} > -0.6$ [**Fig. 5(e)**]. For $\gamma = 0.3$ and $\tilde{\beta}_{13} = 1$, a wider region of G$_2$ phase exists at $x < x_C$; it is followed by a relatively thin region of G$_{123}$ phase, surrounded by G$_{13}$ phase, that is followed by a wider curved region of G$_3$ phase at $x > 0.7$ [**Fig. 5(f)**].

The dependence of the $\tilde{\eta}_\perp$ (red curve) and $\tilde{\eta}_3$ (blue curve) on $x$, calculated at $\gamma = 3$, small $\tilde{\beta}_{13} = +0.1$ and $\tilde{\beta}_{15} = \pm 0.1$ are shown in **Fig. 6(d)**. The $\tilde{\eta}_3$-curve is smooth corresponding to the diffuse second phase transition points. The situation shown in the figure is typical for very small $|\tilde{\beta}_{13}|$.

The amplitude $\tilde{\eta}_\perp$ of the in-plane order parameter components, $\tilde{\eta}_1 = \tilde{\eta}_\perp \cos(\varphi_m)$ and $\tilde{\eta}_2 = \tilde{\eta}_\perp \sin(\varphi_m)$, and the positive $\tilde{\eta}_3$ are shown in **Figs. 7(c-d)**. The dependence of the $\tilde{\eta}_i$ on $x$, calculated for $\gamma = 0.3$, very small $\tilde{\beta}_{13} = +0.1$ and small $\tilde{\beta}_{15} = \pm 0.3$; high $\tilde{\beta}_{13} = +1$ and very small $\tilde{\beta}_{15} = \pm 0.1$ are shown in **Fig. 7(c)** and **7(d)**, respectively. The small features on the $\tilde{\eta}_i$ curves correspond to the phase transition between $G_2$ and $G_{123}$ phases. The appearance of $\tilde{\eta}_3$ at $x \approx (0.5 - 0.6)$ and disappearance of $\tilde{\eta}_\perp$ at $x \approx (0.65 - 0.95)$ corresponds the second order phase transitions, which become diffuse at small $0 < |\tilde{\beta}_{15}| \ll 1$.

Corresponding angles $\varphi_m$ are shown in **Figs. 7(a)** for small $\tilde{\beta}_{13} = 0.1$ and in **Fig.7(b)** for high $\tilde{\beta}_{13} = 1$. There are equidistant 6 angles $\varphi_m$ in $G_2$ phase, which split into 12 angles $\varphi_m$ is $G_{123}$ phase, then merge into 6 angles in $G_{13}$ phase (which are rotated on 30 degree with respect to initial 6 angles), and eventually disappear in $G_3$ phase. Comparing **Fig. 7(a)** and **7(b)**, it is seen that the regions of $G_{123}$ and $G_{13}$ phases stability decrease significantly with $\tilde{\beta}_{13}$ increase from 0.1 to 1.

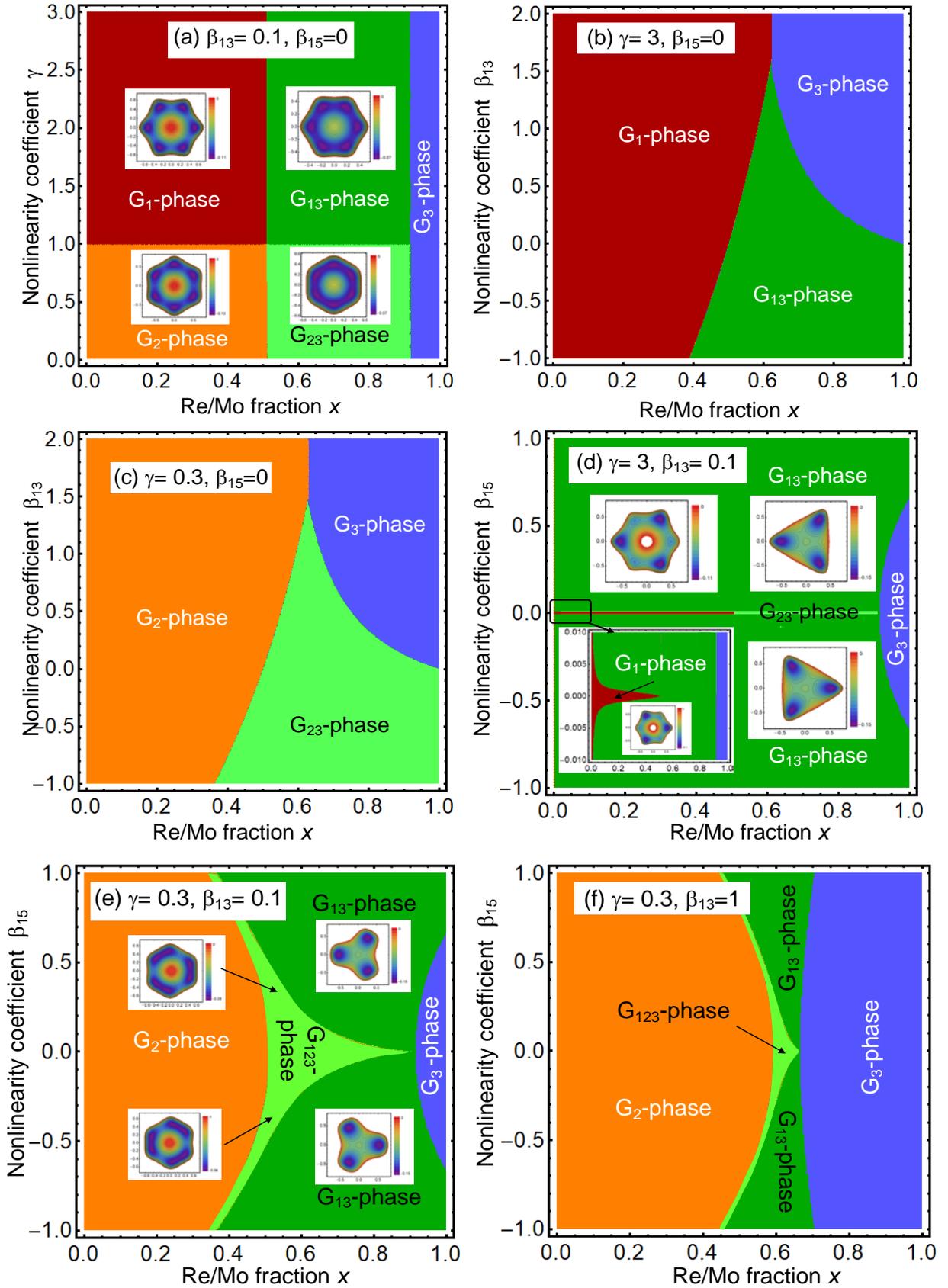

**FIGURE 5.** Phase diagrams of SL-$(MoS_2)_{1-x}$-$(ReS_2)_x$ in dependence of the Re/Mo fraction $x$ and nonlinearity coefficients $\gamma$ (a), $\tilde{\beta}_{13}$ (b,c) and $\tilde{\beta}_{15}$ (e-f). Plots are calculated for $\beta = +1$, $\lambda = +1$, $\delta = 0.5$, $x_{Mo} = 1$, $x_{Re} = 0.5$ and $\Delta = 0.1$, the values of dimensionless parameters $\tilde{\beta}_{13}$, $\tilde{\beta}_{15}$ and $\gamma$ are indicted in the plot legends. Color images in the phase regions schematically show the distribution of the free energy minima in the plane $\{\tilde{\eta}_1, \tilde{\eta}_2\}$ at fixed $\tilde{\eta}_3$ corresponding to the energy minima.

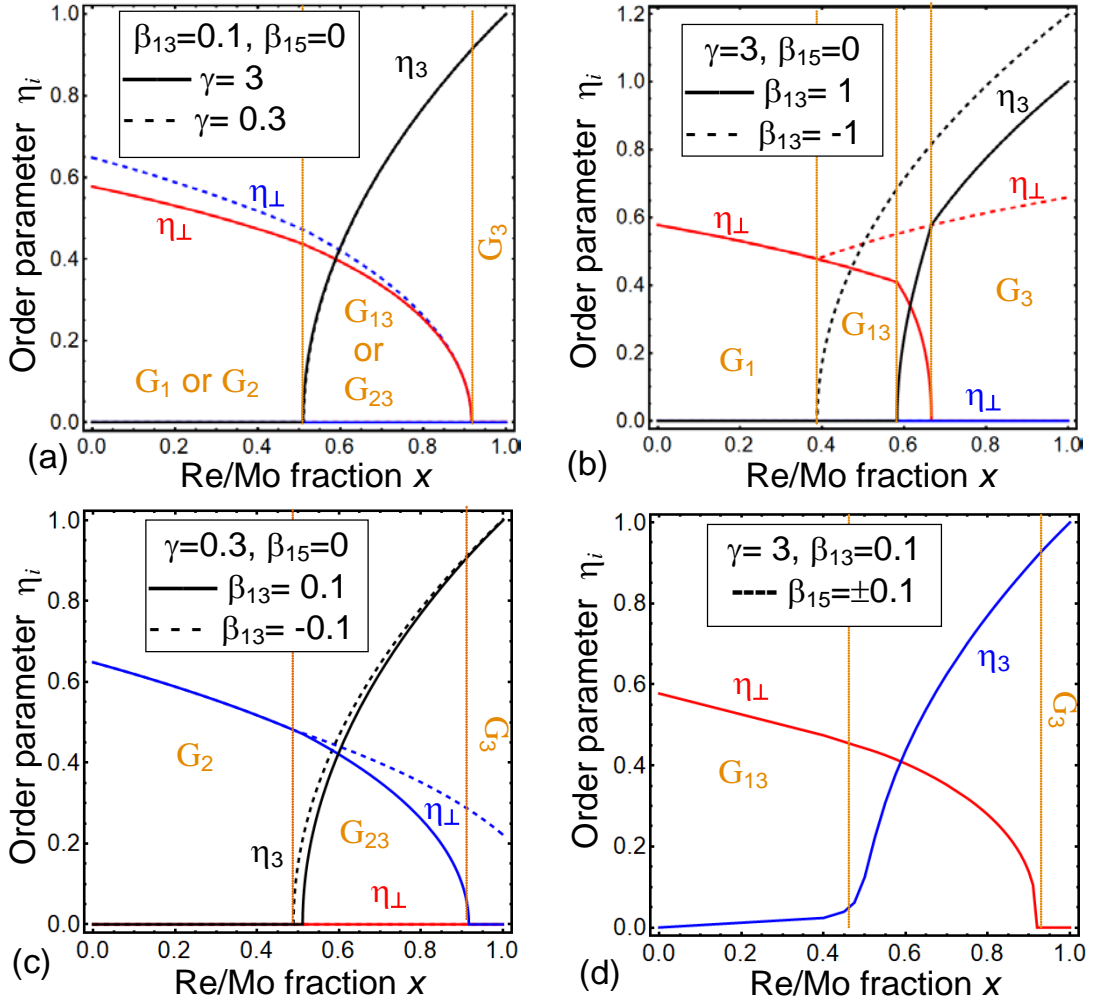

**FIGURE 6.** Values of the order parameter in-plane amplitude $\tilde{\eta}_\perp$ and out-of-plane component $\tilde{\eta}_3$ in dependence on Re/Mo fraction "$x$" in SL-$(MoS_2)_{1-x}$-$(ReS_2)_x$ system. Parameters are the same as in **Fig. 5.**

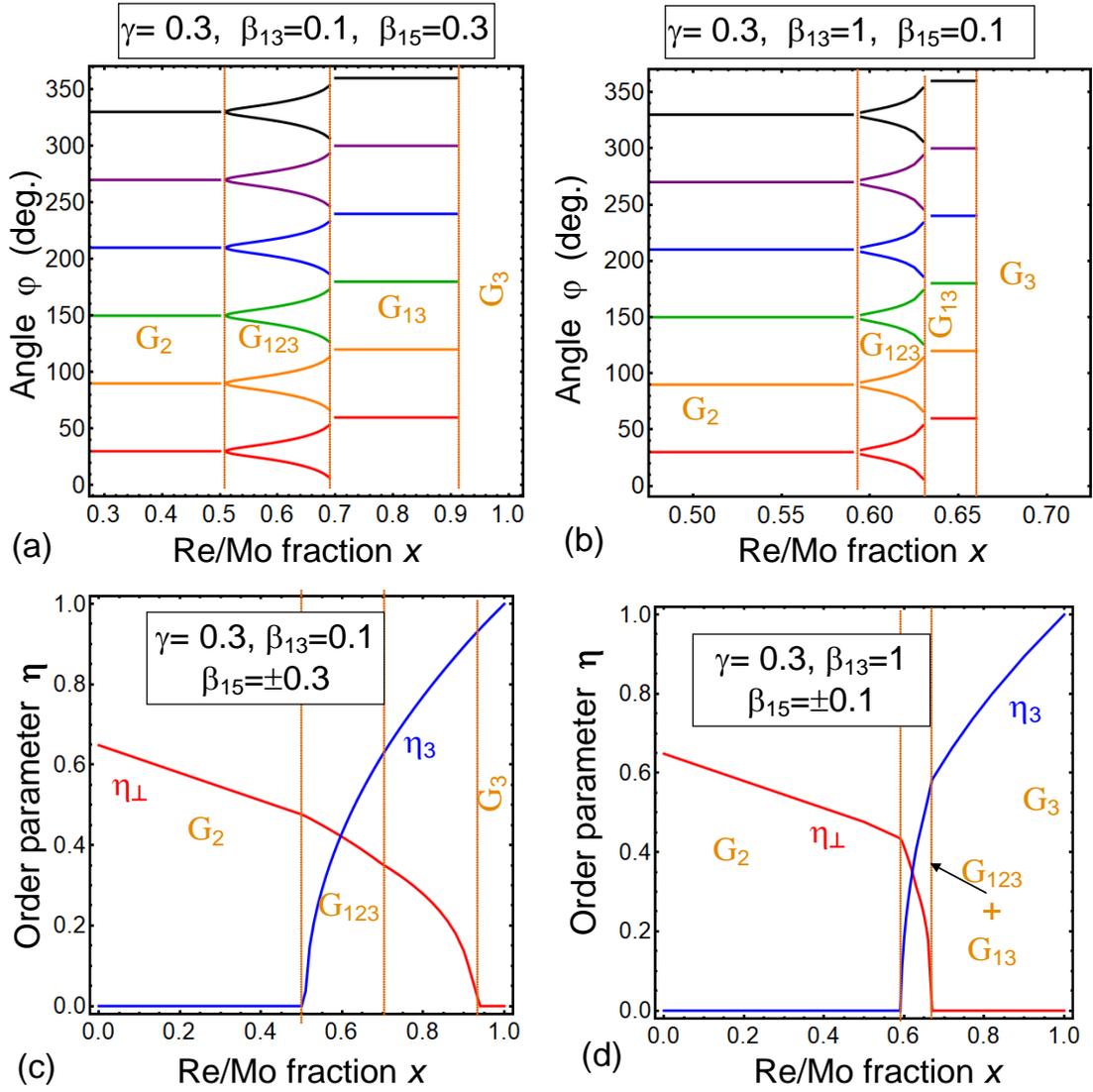

**FIGURE 7.** Values of the order parameter angles φ in plane $\{\tilde{\eta}_1, \tilde{\eta}_2\}$ **(a, b)**; in-plane amplitude $\tilde{\eta}_\perp$ and out-of-plane component $\tilde{\eta}_3$ **(c, d)** in dependence on Re/Mo fraction "*x*" in a SL-$(MoS_2)_{1-x}$-$(ReS_2)_x$. Parameters are the same as in **Fig. 5**.

## V. DISCUSSION

### A. Out-of-plane piezoelectricity and ferroelectricity

The analysis performed in **Section IV** predicts the existence of at least two possible bistable directions of the component $\tilde{\eta}_3$ exists in $G_{13}$, $G_{23}$, $G_{123}$ and $G_3$ structural phases of SL-$(MoS_2)_{1-x}$-$(ReS_2)_x$ [see e.g. **Table II** and **Fig. 8**]. Since $\tilde{\eta}_3$ is polar being related with reflection symmetry breaking normal to the SL plane, the out-of-plane piezoelectricity in $G_{13}$, $G_{23}$, $G_{123}$ and $G_3$ can exist in the structural phases. Since $\tilde{\eta}_3$ is coupled to the polar displacement in the SL, it can be switched between the directions by e.g. external electric field. Note that several stable states in the plane $\{\tilde{\eta}_1, \tilde{\eta}_2\}$ correspond to each direction of $\tilde{\eta}_3$, due to the possibility of free energy minima in-plane rotation at 30 (or 60) degrees. In other words, $\tilde{\eta}_3$ and $\tilde{\eta}_{1,2}$ are coupled in $G_{13}$ and $G_{123}$ (meaning if $\tilde{\eta}_3$ switches, then in plane $\{\tilde{\eta}_1, \tilde{\eta}_2\}$ it also switches by 30 (or 60) degrees. Trigger-type direct transitions from $G_1$, $G_2$ to $G_3$, or from $G_2$ to $G_{13}$ is rare, but not impossible for the considered model [see **Fig.5(b)**].

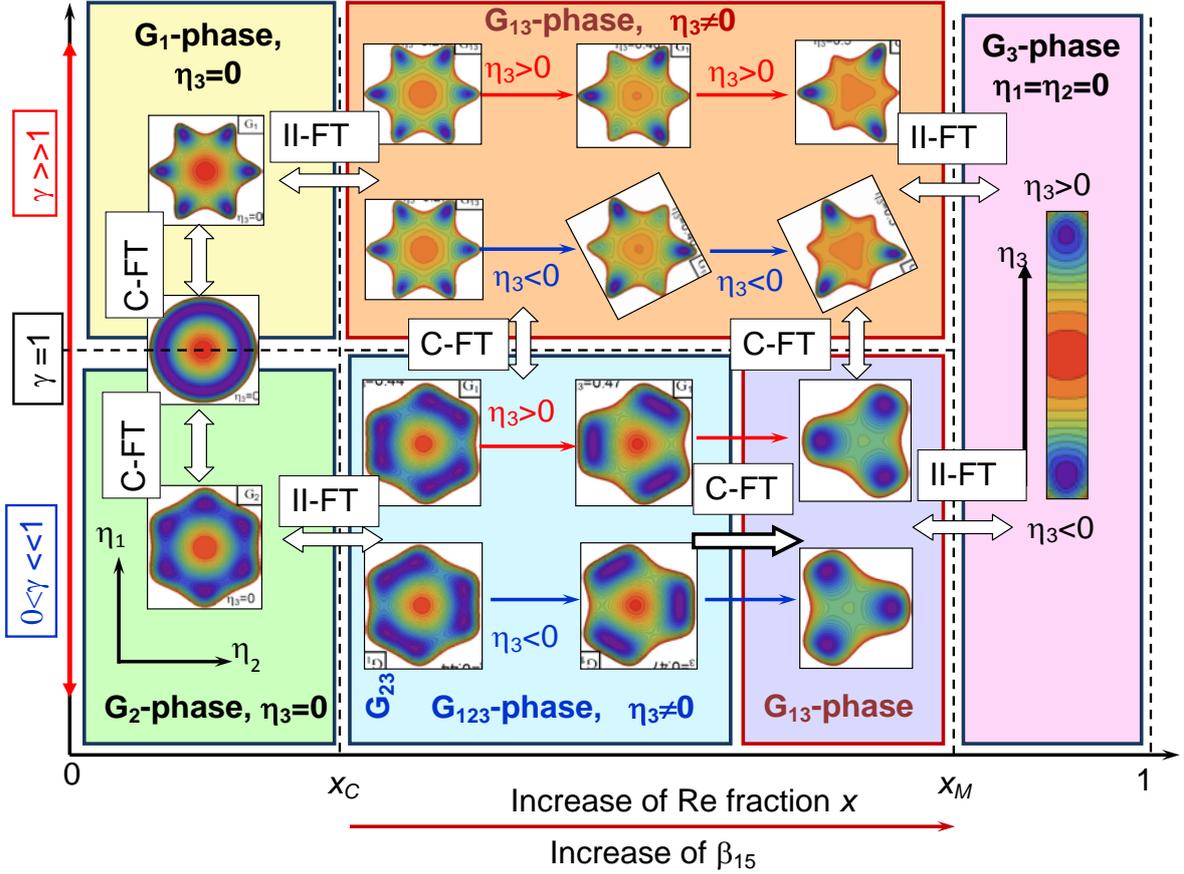

**FIGURE 8.** Schematics of possible phase transformations in a SL-$(MoS_2)_{1-x}$-$(ReS_2)_x$. Color images in the $G_2$, $G_1$, $G_{23}$, $G_{123}$ and $G_{13}$ phases schematically show the evolution of the free energy minima in the plane $\{\tilde{\eta}_1, \tilde{\eta}_2\}$ with x increase. Color bar in $G_3$ phase shows the two minima of $\tilde{\eta}_3$. Abbreviation "II-FT" is for the second order phase transition from the states with $\tilde{\eta}_3 = 0$ ($G_1$ or $G_2$) to the 2 (polar) states with $\tilde{\eta}_3 \neq 0$ ($G_{13}$, or $G_{23}$). II-FT takes place between $G_{13}$ and $G_3$ phases. Abbreviation "C-FT" is for continuous phase transition between e.g. $G_1$ and $G_2$ phases, as well as between $G_{23}$ and $G_{123}$; $G_{13}$ and $G_{123}$, and different configurations of $G_{13}$ phases.

Thus, the out-of-plane and in-plane polar displacements can be coupled similarly to the bulk multiaxial ferroelectrics. Obtained numerical and analytical results allow us to estimate the values of spontaneous polarization assuming that its components are proportional to the vector $\{\tilde{\eta}_1, \tilde{\eta}_2, \tilde{\eta}_3\}$, and predict the existence of 2, 6 or 12 polar domain types in different structural phases. These results are in agreement with recent works [32, 33, 34, 35, 36, 37].

Since different structural phases of SL-TMDs can be either conducting or semiconducting [5, 6], different domains can be distinguished by conducting properties. However, the domain engineering is not the only possibility to tune the SL-TMDs conductivity. Complementary, one should take into consideration the strain concentration (and possible band changes) at the domain walls, which can be rather thin and attracting as conductivity channels.

Above calculations are performed assuming that the spatial gradients of the order parameter components are absent. The assumption can be valid in a model case of mechanically free and flat infinite SL. Any bending of suspended SL can induce the gradient of $\tilde{\eta}_i$. Due to the flexoelectric coupling it immediately leads to the polarization changes. The impact of the flexoelectric coupling on the phase diagrams and structural properties of SL-$(MoS_2)_{1-x}$-$(ReS_2)_x$ will be studied in our forthcoming papers. However, the important sequence of the gradient-related elastic strain is that the inhomogeneous strain in the vicinity of SL boundaries should induce the changes of the SL conductivity due to e.g. the coupling with band structure via deformation potential.

The deformation potential value for $MoS_2$ (band gap about 1.7 eV) can be estimated from Johari and Shenoy *ab initio* results [4] as $\Sigma = 17$ eV. The high enough value is very close to the graphene one, and so it is quite realistic. The conductivity change at the wall is proportional to $\exp\left(-\frac{\Sigma_{ij} u_{ij}}{k_B T}\right)$ (Boltzmann approximation for non-degenerated carriers) and $u_{ij}[\boldsymbol{\eta}] - u_{ij}[0] \approx (\nabla u_{ij}[0])\boldsymbol{\eta}$. Thus we can expect that all types of domain walls in SL-$(MoS_2)_{1-x}$-$(ReS_2)_x$ should become conductive, once the spontaneous strain value exceeds several %.

## VI. SUMMARY

We derived the Landau-type free energy functional for SL-$(MoS_2/ReS_2)$ and analyze the free energy relief, phase diagrams and order parameter behavior of SL-$(MoS_2)_{1-x}$-$(ReS_2)_x$. Our results predict the existence of multiple structural phases with 2-, 6- and 12-fold degenerated energy minima. The out-of-plane piezoelectricity and ferroelectricity can exist in many of these phases, regarding that value of switchable spontaneous polarization is proportional to the out-of-plane order parameter, and predict the existence of 2, 6 or 12 types of polar domain in different phases. Also we can expect that the domain walls in SL-$(MoS_2)_{1-x}$-$(ReS_2)_x$ should become conductive, as the spontaneous strain value exceeds several %.

Our analysis (being limited to the point symmetry) can be generalized for other 2D materials immobilized in 2D configurations. We further note that for free standing materials further interesting set of phenomena should emerge due to bucking in the direction normal to SL, when the primary flexoelectricty of 2D system will start playing a role [38].

**Acknowledgements.** This material is based upon work (S.V.K, R.V.) supported by the U.S. Department of Energy, Office of Science, Office of Basic Energy Sciences, and performed in the Center for Nanophase Materials Sciences, supported by the Division of Scientific User Facilities. A portion of FEM was conducted at the Center for Nanophase Materials Sciences, which is a DOE Office of Science User Facility (CNMS Proposal ID: 257). This work (Y.K.) was supported by "Human Resources Program in Energy Technology" of the Korea Institute of Energy Technology Evaluation and Planning (KETEP), granted financial resource from the Ministry of Trade, Industry & Energy, Republic of Korea. (No. 20174030201800) A.N.M work supported by the National Academy of Sciences of Ukraine and has received funding from the European Union's Horizon 2020 research and innovation programme under the Marie Skłodowska-Curie grant agreement No 778070.

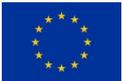

**Authors' contribution.** A.N.M. evolved the theoretical description, performed analytical calculations jointly with E.A.E., and interpreted numerical results, obtained by E.A.E. and K.S. S.V.K. generated the research idea and, jointly with A.N.M., wrote the manuscript draft. R.V. and Y.K. worked on the results discussion and manuscript improvement.

## Appendix A.

Direct application of aristo-phase point group symmetry operations to the tensors $\alpha_{ij}$, $\beta_{ijkl}$, $\gamma_{ijklmn}$, $g_{ijkl}$, $Q_{ijkl}$ and $R_{ijklmn}$ along with existing index-permutation symmetry, leads to the system of linear algebraic equations for each tensor:

$$\alpha_{ij} = C_{ii'}C_{jj'}\alpha_{i'j'}, \quad \begin{cases} \beta_{ijkl} = C_{ii'}C_{jj'}C_{kk'}C_{ll'}\beta_{i'j'k'l'}, \\ \beta_{ijkl} = \beta_{jikl} = \beta_{jilk} = \beta_{klij} = ... \end{cases}, \quad g_{ijkl} = C_{ii'}C_{jj'}C_{kk'}C_{ll'}g_{i'j'k'l'} \quad (A.1)$$

$$\begin{cases} \gamma_{ijklmn} = C_{ii'}C_{jj'}C_{kk'}C_{ll'}C_{mm'}C_{nn'}\gamma_{i'j'k'l'm'n'}, \\ \gamma_{ijklmn} = \gamma_{jiklmn} = \gamma_{jkilmn} = \gamma_{kijlmn} = \gamma_{ikjlmn} = \gamma_{kjilmn} = ... \end{cases}, \quad (A.2)$$

$$\begin{cases} c_{ijkl} = C_{ii'}C_{jj'}C_{kk'}C_{ll'}c_{i'j'k'l'}, \\ c_{ijkl} = c_{jikl}, \quad c_{ijkl} = c_{ijlk}, \quad c_{ijkl} = c_{klij} \end{cases}, \quad \begin{cases} Q_{ijkl} = C_{ii'}C_{jj'}C_{kk'}C_{ll'}Q_{i'j'k'l'}, \\ Q_{ijkl} = Q_{jikl}, \quad Q_{ijkl} = Q_{ijlk}. \end{cases}, \quad (A.3)$$

$$\begin{cases} R_{ijklmn} = C_{ii'}C_{jj'}C_{kk'}C_{ll'}C_{mm'}C_{nn'}R_{i'j'k'l'm'n'}, \\ R_{ijklmn} = R_{jiklmn}, \quad R_{ijklmn} = R_{ijlkmn} = R_{ijlknm} = R_{ijmnkl} = .... \end{cases} \quad (A.4)$$

The first lines in Eqs.(A.1)-(A.4) include point group symmetry, where matrix elements $C_{ii'}$ are $\bar{6}m2$ (for a bulk), $6mm$ or $3m$ (for a SL) group symmetry operations in the matrix form, and the second line reflects the index-permutation symmetry (if any exist). Solution of Eqs.(A.1)-(A.4) gives the detailed structure of the tensors, including the number of nonzero and independent elements. Namely for $6mm$ we obtained that

$$\hat{\alpha} = \begin{pmatrix} \alpha_{11} & 0 & 0 \\ 0 & \alpha_{11} & 0 \\ 0 & 0 & \alpha_{33} \end{pmatrix} \quad (A.5a)$$

$$\hat{\beta} = \begin{pmatrix} \beta_{11} & \beta_{11} & \beta_{13} \\ \beta_{11} & \beta_{11} & \beta_{13} \\ \beta_{13} & \beta_{13} & \beta_{33} \end{pmatrix} \quad (A.5b)$$

Here we introduced Voight notations $\beta_{11} \equiv \beta_{1111}$, $\beta_{13} \equiv 6\beta_{1133}$, $\beta_{33} \equiv \beta_{3333}$ and used the relation $\beta_{1122} = \dfrac{\beta_{1111}}{3}$. Elastic compliances and striction tensors are

$$\hat{c} = \begin{pmatrix} c_{1111} & c_{1122} & c_{1133} & 0 & 0 & 0 \\ c_{1122} & c_{1111} & c_{1133} & 0 & 0 & 0 \\ c_{1133} & c_{1133} & c_{3333} & 0 & 0 & 0 \\ 0 & 0 & 0 & c_{1313} & 0 & 0 \\ 0 & 0 & 0 & 0 & c_{1313} & 0 \\ 0 & 0 & 0 & 0 & 0 & c_{1212} \end{pmatrix}, \quad c_{1212} = \frac{c_{1111} - c_{1122}}{2} \tag{A.6}$$

$$\hat{Q} = \begin{pmatrix} Q_{1111} & Q_{1122} & Q_{1133} & 0 & 0 & 0 \\ Q_{1122} & Q_{1111} & Q_{1133} & 0 & 0 & 0 \\ Q_{3311} & Q_{3311} & Q_{3333} & 0 & 0 & 0 \\ 0 & 0 & 0 & Q_{1313} & 0 & 0 \\ 0 & 0 & 0 & 0 & Q_{1313} & 0 \\ 0 & 0 & 0 & 0 & 0 & Q_{1212} \end{pmatrix} \quad Q_{1212} = \frac{Q_{1111} - Q_{1122}}{2} \tag{A.7}$$

The 2-nd and 4-th order convolutions are:

$$\alpha_{ij}\eta_i\eta_j = \alpha_{11}\left(\eta_1^2 + \eta_2^2\right) + \alpha_{33}\eta_3^2, \tag{A.8a}$$

$$\beta_{ijkl}\eta_i\eta_j\eta_k\eta_l = \beta_{11}\left(\eta_1^2 + \eta_2^2\right)^2 + \beta_{13}\left(\eta_1^2 + \eta_2^2\right)\eta_3^2 + \beta_{33}\eta_3^4, \tag{A.8b}$$

Here we introduced Voight notations $\beta_{11} \equiv \beta_{1111}$, $\beta_{13} \equiv 6\beta_{1133}$, $\beta_{33} \equiv \beta_{3333}$ and used the relation $\beta_{1122} = \frac{\beta_{1111}}{3}$. Equations (A.8a-b) have 5 independent and unknown constants for 3D case with $\eta_3 \neq 0$, and 2 constants for a 2D case with $\eta_3 = 0$.

The 6-th order convolution is:

$$\begin{aligned}\gamma_{ijklmn}\eta_i\eta_j\eta_k\eta_l\eta_m\eta_n &= \gamma_{111}\eta_1^2\left(\eta_1^2 - 3\eta_2^2\right)^2 + \gamma_{222}\eta_2^2\left(3\eta_1^2 - \eta_2^2\right)^2 \\ &+ \gamma_{113}\left(\eta_1^2 + \eta_2^2\right)^2\eta_3^2 + \gamma_{133}\left(\eta_1^2 + \eta_2^2\right)\eta_3^4 + \gamma_{333}\eta_3^6\end{aligned}. \tag{A.8c}$$

Here we introduced $\gamma_{111} \equiv \gamma_{111111}$, $\gamma_{222} \equiv \gamma_{222222}$, $\gamma_{113} \equiv 15\gamma_{111133}$, $\gamma_{133} \equiv 15\gamma_{113333}$ and $\gamma_{333} \equiv \gamma_{333333}$. Equation (A.8c) has 5 independent constants for 3D case with $\eta_3 \neq 0$, and 2 constants for a 2D case with $\eta_3 = 0$.

**Appendix B. Analytical solutions**

One obtains from Eq.(5) three coupled equations of state for the components of the order parameter $\mathbf{\eta} = (\eta_1, \eta_2, \eta_3)$:

$$\begin{cases}
2\alpha_{11}\eta_1 + 4\beta_{11}(\eta_1^2 + \eta_2^2)\eta_1 + 2\beta_{13}\eta_3^2\eta_1 + 3\beta_{15}\eta_3(\eta_1^2 - \eta_2^2) + 6\gamma_{111}(\eta_1^4 - 4\eta_1^2\eta_2^2 + 3\eta_2^4)\eta_1 + 2\gamma_{133}\eta_3^4\eta_1 \\
+ 12\gamma_{222}(3\eta_1^2 - \eta_2^2)\eta_2^2\eta_1 + 4\gamma_{113}(\eta_1^2 + \eta_2^2)\eta_3^2\eta_1 + 6\gamma_{115}\eta_3(5\eta_1^4 - 6\eta_1^2\eta_2^2 - 3\eta_2^4) + 3\gamma_{135}\eta_3^3(\eta_1^2 - \eta_2^2) = 0, \\[4pt]
2\alpha_{11}\eta_2 + 4\beta_{11}(\eta_1^2 + \eta_2^2)\eta_2 + 2\beta_{13}\eta_3^2\eta_2 - 6\beta_{15}\eta_1\eta_2\eta_3 + 6\gamma_{222}(\eta_2^4 - 4\eta_1^2\eta_2^2 + 3\eta_1^4)\eta_2 \\
+ 12\gamma_{111}(3\eta_2^2 - \eta_1^2)\eta_1^2\eta_2 + 4\gamma_{113}(\eta_1^2 + \eta_2^2)\eta_3^2\eta_2 + 2\gamma_{133}\eta_3^4\eta_2 - 24\gamma_{115}\eta_1\eta_2\eta_3(\eta_1^2 + 3\eta_2^2) - 6\gamma_{135}\eta_{11}\eta_2\eta_3^3 = 0, \\[4pt]
2\alpha_{33}\eta_3 + 2\beta_{13}(\eta_1^2 + \eta_2^2)\eta_3 + \beta_{15}\eta_1(\eta_1^2 - 3\eta_2^2) + 4\beta_{33}\eta_3^3 + \\
2\gamma_{113}(\eta_1^2 + \eta_2^2)^2\eta_3 + 4\gamma_{133}(\eta_1^2 + \eta_2^2)\eta_3^3 + 6\gamma_{333}\eta_3^5 + \gamma_{115}\eta_1(\eta_1^2 - 3\eta_2^2)\eta_\perp^2 + 3\gamma_{135}\eta_1\eta_3^2(\eta_1^2 - 3\eta_2^2) = 0.
\end{cases}$$
(B.1)

Possible stable phases of free energy (5) corresponding to solutions of Eq.(B.1) for $\eta_3 = 0$ are listed in **Table SI**.

**Table SI.** Possible stable phases of free energy (5) valid for $\eta_3 = 0$

| Phase | Order parameter components $\eta_1$, $\eta_2$ | Phase energy and stability conditions |
|---|---|---|
| Disordered | $\eta_1 = \eta_2 = 0$, $\eta_\perp = \sqrt{\eta_1^2 + \eta_2^2} = 0$ | $f_L^0 = 0 = \min$. Stability conditions are $\beta_{11} > 0$, $\alpha_{11} \geq 0$ or $\beta_{11} < 0$, $\gamma_{111} \geq 0$, $\gamma_{222} \geq 0$, $\beta_{11}^2 - 4\alpha_{11}\gamma_{222} \leq 0$, $\beta_{11}^2 - 4\alpha_{11}\gamma_{111} \leq 0$ |
| $G_2$-phase | $\eta_1 = \eta_\perp \cos(\varphi_m)$ $\eta_2 = \eta_\perp \sin(\varphi_m)$ $\eta_\perp = \sqrt{\dfrac{-\alpha_{11}}{\sqrt{\beta_{11}^2 - 3\alpha_{11}\gamma_{222}} + \beta_{11}}}$ $\varphi_m = \dfrac{\pi}{6} + \dfrac{\pi}{3}m$, $m = 0, ..., 5$. | $f_L^{G2} = \dfrac{-\alpha_{11}^2[\beta_{11}^2 - 4\alpha_{11}\gamma_{222}]}{2[\beta_{11}^2 - 3\alpha_{11}\gamma_{222}]^{3/2} + 2\beta_{11}^3 - 9\alpha_{11}\beta_{11}\gamma_{222}} = \min$ Stability conditions are $\alpha_{11} < 0$, $\gamma_{111} \geq \gamma_{222} \geq 0$ or $\alpha_{11} > 0$, $\beta_{11} < 0$, $\beta_{11}^2 \geq 4\alpha_{11}\gamma_{222}$, $\gamma_{111} \geq \gamma_{222} \geq 0$ |
| $G_1$-phase | $\eta_1 = \eta_\perp \cos(\varphi_m)$ $\eta_2 = \eta_\perp \sin(\varphi_m)$ $\eta_\perp = \sqrt{\dfrac{-\alpha_{11}}{\sqrt{\beta_{11}^2 - 3\alpha_{11}\gamma_{111}} + \beta_{11}}}$ $\varphi_m = \dfrac{\pi}{3}m$, $m = 0, ..., 5$. | $f_L^{G1} = \dfrac{-\alpha_{11}^2[\beta_{11}^2 - 4\alpha_{11}\gamma_{111}]}{2[\beta_{11}^2 - 3\alpha_{11}\gamma_{111}]^{3/2} + 2\beta_{11}^3 - 9\alpha_{11}\beta_{11}\gamma_{111}} = \min$ Stability conditions are $\alpha_{11} < 0$, $\gamma_{222} \geq \gamma_{111} \geq 0$ or $\alpha_{11} > 0$, $\beta_{11} < 0$, $\beta_{11}^2 \geq 4\alpha_{11}\gamma_{111}$, $\gamma_{222} \geq \gamma_{111} \geq 0$ |

Now let us consider more general case when $\eta_3$ can be nonzero. Using polar coordinate system and putting $\tilde{\gamma}_{113} = \tilde{\gamma}_{133} = \tilde{\gamma}_{333} = \tilde{\gamma}_{115} = \tilde{\gamma}_{135} = 0$, one obtains from Eq.(B.1) three coupled equations of state for $\tilde{\eta}_i$ and $\varphi$

$$\begin{cases} (\gamma-1)\tilde{\eta}_\perp^6 \sin(6\varphi) - \tilde{\beta}_{15} b \tilde{\eta}_\perp^3 \tilde{\eta}_3 \sin(3\varphi) \equiv \left[2(\gamma-1)\tilde{\eta}_\perp^6 \cos(3\varphi) - \tilde{\beta}_{15} b \tilde{\eta}_\perp^3 \tilde{\eta}_3\right]\sin(3\varphi) = 0, \\ 2a\tilde{\eta}_\perp + 4\beta\tilde{\eta}_\perp^3 + 6\left[\cos^2(3\varphi) + \gamma\sin^2(3\varphi)\right]\tilde{\eta}_\perp^5 + b\left[2\tilde{\beta}_{13}\tilde{\eta}_3^2\tilde{\eta}_\perp + 3\tilde{\beta}_{15}\tilde{\eta}_\perp^2\tilde{\eta}_3\cos(3\varphi)\right] = 0, \\ 2\left(\lambda c + \tilde{\beta}_{13} b \tilde{\eta}_\perp^2\right)\tilde{\eta}_3 + 4\delta\tilde{\eta}_3^3 + \tilde{\beta}_{15} b \tilde{\eta}_\perp^3 \cos(3\varphi) = 0. \end{cases} \quad (B.2)$$

From these equations the angles can be found from equations, $\cos(3\varphi) = \dfrac{\tilde{\beta}_{15} b \tilde{\eta}_3}{2(\gamma-1)\tilde{\eta}_\perp^3}$ (case 1) or $\sin(3\varphi) = 0$ (case 2). For these 2 cases Eqs.(B.2) split into the systems:

**Case 1.** For $\cos(3\varphi) = \dfrac{\tilde{\beta}_{15} b \tilde{\eta}_3}{2(\gamma-1)\tilde{\eta}_\perp^3}$ the free energy is

$$\begin{aligned} \tilde{g} &= \begin{pmatrix} a(x)\tilde{\eta}_\perp^2 + \beta\tilde{\eta}_\perp^4 + \left[(1-\gamma)\cos^2(3\varphi) + \gamma\right]\tilde{\eta}_\perp^6 + \\ \lambda c(x)\tilde{\eta}_3^2 + \delta\tilde{\eta}_3^4 + b(x)\tilde{\eta}_\perp^2\left[\tilde{\beta}_{13}\tilde{\eta}_3^2 + \tilde{\beta}_{15}\tilde{\eta}_\perp\tilde{\eta}_3\cos(3\varphi)\right] \end{pmatrix} \\ &\equiv a\tilde{\eta}_\perp^2 + \beta\tilde{\eta}_\perp^4 + \gamma\tilde{\eta}_\perp^6 + \left(\lambda c + \dfrac{\tilde{\beta}_{15}^2 b^2}{4(\gamma-1)}\right)\tilde{\eta}_3^2 + \delta\tilde{\eta}_3^4 + \tilde{\beta}_{13} b \tilde{\eta}_\perp^2 \tilde{\eta}_3^2 \end{aligned} \quad (B.3)$$

Minimization of the energy (B.3) with respect to $\tilde{\eta}_3$ and $\tilde{\eta}_\perp$ leads to the equations:

$$\begin{cases} \left(\lambda c + \dfrac{\tilde{\beta}_{15}^2 b^2}{4(\gamma-1)} + \tilde{\beta}_{13} b \tilde{\eta}_\perp^2\right)\tilde{\eta}_3 + 2\delta\tilde{\eta}_3^3 = 0, \Rightarrow \tilde{\eta}_3 = 0 \text{ or } \tilde{\eta}_3^2 = -\dfrac{\lambda c + \tilde{\beta}_{13} b \tilde{\eta}_\perp^2}{2\delta} - \dfrac{\tilde{\beta}_{15}^2 b^2}{4\delta(\gamma-1)}, \\ \tilde{\eta}_\perp\left(a + \tilde{\beta}_{13} b \tilde{\eta}_3^2 + 2\beta\tilde{\eta}_\perp^2 + 3\gamma\tilde{\eta}_\perp^4\right) = 0, \Rightarrow \\ \Rightarrow \tilde{\eta}_\perp\left(a - \lambda c \dfrac{\tilde{\beta}_{13} b}{2\delta} - \dfrac{\tilde{\beta}_{13}\tilde{\beta}_{15}^2 b^3}{4\delta(\gamma-1)} + \left(2\beta - \dfrac{\tilde{\beta}_{13}^2 b^2}{2\delta}\right)\tilde{\eta}_\perp^2 + 3\gamma\tilde{\eta}_\perp^4\right) = 0, \text{ or } \tilde{\eta}_\perp\left(a + 2\beta\tilde{\eta}_\perp^2 + 3\gamma\tilde{\eta}_\perp^4\right) = 0 \Rightarrow \end{cases}$$

(B.4)

There are 2 groups of solutions of Eq.(B.4), namely:

$$\begin{cases} \tilde{\eta}_\perp = \sqrt{\dfrac{1}{3\gamma}\left(\sqrt{\beta^{*2} - 3\gamma a^*} - \beta^*\right)}, \\ \tilde{\eta}_3 = \pm\sqrt{-\dfrac{\lambda c + \tilde{\beta}_{13} b \tilde{\eta}_\perp^2}{2\delta} - \dfrac{\tilde{\beta}_{15}^2 b^2}{4\delta(\gamma-1)}}. \end{cases} \quad (B.5a)$$

where $a^* = a - \lambda c \dfrac{\tilde{\beta}_{13} b}{2\delta} - \dfrac{\tilde{\beta}_{13}\tilde{\beta}_{15}^2 b^3}{4\delta(\gamma-1)}$ and $\beta^* = \beta - \dfrac{\tilde{\beta}_{13}^2 b^2}{4\delta}$. Since $\tilde{\beta}_{15} b \tilde{\eta}_3 \tilde{\eta}_\perp^3 \cos(3\varphi) \equiv \dfrac{\tilde{\beta}_{15}^2 b^2 \tilde{\eta}_3^2}{2\delta(\gamma-1)} = \min$ for

the solutions (B.5a), they are "mixed" G123 phase. The polar angle has 6 directions

$$\cos(3\varphi) = \dfrac{\tilde{\beta}_{15} b \tilde{\eta}_3}{2(\gamma-1)\tilde{\eta}_\perp^3}, \quad \varphi = \pm\dfrac{1}{3}\arccos\left[\dfrac{\tilde{\beta}_{15} b \tilde{\eta}_3}{2(\gamma-1)\tilde{\eta}_\perp^3}\right] + \dfrac{2\pi m}{3}, \quad m = 0, 1, 2. \quad (B.5b)$$

The 6-fold degenerated energy value corresponding to extremum is

$$\tilde{g}_L^{ext} = \frac{-a^{*2}(\beta^{*2} - 4\gamma a^*)}{2(\beta^{*2} - 3\gamma a^*)^{3/2} + 2\beta^{*3} - 9\gamma\beta^* a^*}, \qquad (B.5c)$$

The energy extremum is minimum at $\gamma < 1$.

Another solution is:

$$\tilde{\eta}_\perp^2 = \frac{1}{3\gamma}\left(\sqrt{\beta^2 - 3\gamma a} - \beta\right) \equiv \frac{-a}{\sqrt{\beta^2 - 3\gamma a} + \beta}, \quad \tilde{\eta}_3 = 0, \quad \varphi = \pm\frac{\pi}{6} + \frac{2\pi m}{3} \qquad (B.5d)$$

The energy of the solution (B.5d) is already listed in **Table III** as "pure" G2 phases.

**Case 2.** For $\sin(3\varphi) = 0$, $\varphi = \frac{\pi m}{3}$, $m = 0,1,..5.$, the free energy is

$$\tilde{g} = a\tilde{\eta}_\perp^2 + \beta\tilde{\eta}_\perp^4 + \tilde{\eta}_\perp^6 + \lambda c\tilde{\eta}_3^2 + \delta\tilde{\eta}_3^4 + b\tilde{\eta}_\perp^2\left[\tilde{\beta}_{13}\tilde{\eta}_3^2 + (-1)^m \tilde{\beta}_{15}\tilde{\eta}_\perp\tilde{\eta}_3\right] \qquad (B.6)$$

Minimization of the energy (B.6) with respect to $\tilde{\eta}_\perp$ and $\tilde{\eta}_3$ leads to the equations:

$$\begin{cases} \tilde{\eta}_\perp\left(2a + 4\beta\tilde{\eta}_\perp^2 + 6\tilde{\eta}_\perp^4 + b\left[2\tilde{\beta}_{13}\tilde{\eta}_3^2 + (-1)^m 3\tilde{\beta}_{15}\tilde{\eta}_\perp\tilde{\eta}_3\right]\right) = 0, \\ 2(\lambda c + \tilde{\beta}_{13}b\tilde{\eta}_\perp^2)\tilde{\eta}_3 + 4\delta\tilde{\eta}_3^3 + (-1)^m \tilde{\beta}_{15}b\tilde{\eta}_\perp^3 = 0. \end{cases} \qquad (B.7)$$

Assuming that the appearance of $\tilde{\eta}_3$ obeys the second order phase transition scenario, while the appearance of $\tilde{\eta}_\perp$ can be of the first order, approximate solution of Eqs.(B.7) becomes possible for small enough $|\tilde{\beta}_{15}b| \ll 1$ in the vicinity of the $\tilde{\eta}_3$ second order phase transition. Here $\tilde{\eta}_3$ is too small to affect $\tilde{\eta}_\perp$ strongly and so in the first approximation the solutions of the second Eq.(B.7) are

$$\tilde{\eta}_3 \approx \begin{cases} \pm\sqrt{-\dfrac{\lambda c + \tilde{\beta}_{13}b\tilde{\eta}_\perp^2}{2\delta}} + \dfrac{(-1)^m \tilde{\beta}_{15}b\tilde{\eta}_\perp^3}{4(\lambda c + \tilde{\beta}_{13}b\tilde{\eta}_\perp^2)}, & \lambda c + \tilde{\beta}_{13}b\tilde{\eta}_\perp^2 < 0, \\ -\dfrac{(-1)^m \tilde{\beta}_{15}b\tilde{\eta}_\perp^3}{2(\lambda c + \tilde{\beta}_{13}b\tilde{\eta}_\perp^2)}, & \lambda c + \tilde{\beta}_{13}b\tilde{\eta}_\perp^2 > 0 \end{cases} \qquad (B.8)$$

The perturbation approach can be used for small $|\tilde{\beta}_{13}b| \ll 1$ and $|\tilde{\beta}_{15}b| \ll 1$, which become small simultaneously.

**2a)** Let us consider $\lambda c + \tilde{\beta}_{13}b\tilde{\eta}_\perp^2 < 0$ at first. The in-plane components amplitude allowing for Eq.(B.7)-(B.8) are

$$\begin{bmatrix} \tilde{\eta}_\perp = 0 \\ 2a - \tilde{\beta}_{13}b\dfrac{\lambda c}{\delta} + 4\left(\beta - \dfrac{\tilde{\beta}_{13}^2 b^2}{4\delta}\right)\tilde{\eta}_\perp^2 + 6\tilde{\eta}_\perp^4 \approx -(-1)^m 3\tilde{\beta}_{15}b\tilde{\eta}_\perp\sqrt{-\dfrac{\lambda c}{2\delta}} \end{bmatrix} \qquad (B.9a)$$

At $\beta_{15} = 0$ the last equation has the solution $\tilde{\eta}_\perp = \sqrt{\frac{1}{3}\left(\sqrt{\beta^{*2} - 3a^*} - \beta^*\right)}$, where $a^* = a - \lambda c \frac{\tilde{\beta}_{13}b}{2\delta}$ and $\beta^* = \beta - \frac{\tilde{\beta}_{13}^2 b^2}{4\delta}$. The solution has 12-fold degenerated energy extremums

$$\tilde{g}_L^{ext} = \frac{-a^{*2}(\beta^{*2} - 4a^*)}{2(\beta^{*2} - 3a^*)^{3/2} + 2\beta^{*3} - 9\beta^* a^*}.$$ For $0 < |\tilde{\beta}_{15}b| \ll 1$ the energy value can be approximated as

$$\begin{aligned}
\tilde{g}_L^{ext} &\approx \frac{-a^{*2}(\beta^{*2} - 4a^*)}{2(\beta^{*2} - 3a^*)^{3/2} + 2\beta^{*3} - 9\beta^* a^*} + (-1)^m \tilde{\beta}_{15} b \tilde{\eta}_\perp^3 \tilde{\eta}_3 \\
&\approx \frac{-a^{*2}(\beta^{*2} - 4a^*)}{2(\beta^{*2} - 3a^*)^{3/2} + 2\beta^{*3} - 9\beta^* a^*} \pm (-1)^m \frac{\tilde{\beta}_{15}b}{3\sqrt{3}}\left(\sqrt{\beta^{*2} - 3a^*} - \beta^*\right)^{3/2} \sqrt{-\frac{\lambda c}{2\delta}}
\end{aligned}$$ (B.9b)

Expression (B.9b) is 6-fold degenerated for the each sign of $\tilde{\eta}_3$ (positive or negative), indicating the transition of a planar hexagon minima to 2 triangles minima located in different planes $\tilde{\eta}_3 \approx +\sqrt{-\frac{\lambda c}{2\delta}}$ for $(-1)^m \tilde{\beta}_{15}b < 0$ and $\tilde{\eta}_3 \approx -\sqrt{-\frac{\lambda c}{2\delta}}$ for $(-1)^m \tilde{\beta}_{15}b > 0$.

**2b)** Now let us consider $\lambda c + \tilde{\beta}_{13} b \tilde{\eta}_\perp^2 > 0$. The in-plane components amplitude allowing for Eq.(B.7)-(B.8) are

$$\begin{cases} \tilde{\eta}_\perp = 0 \\ 2a + 4\beta\tilde{\eta}_\perp^2 + 6\tilde{\eta}_\perp^4 + \frac{\tilde{\beta}_{13}\tilde{\beta}_{15}^2 b^3 \tilde{\eta}_\perp^6}{2(\lambda c + \tilde{\beta}_{13}b\tilde{\eta}_\perp^2)^2} = \frac{3\tilde{\beta}_{15}^2 b^2 \tilde{\eta}_\perp^4}{2(\lambda c + \tilde{\beta}_{13}b\tilde{\eta}_\perp^2)} \end{cases}$$ (B.10a)

The last equation can be simplified as $a + 2\beta\tilde{\eta}_\perp^2 + 3\left(1 - \frac{\tilde{\beta}_{15}^2 b^2}{4\lambda c}\right)\tilde{\eta}_\perp^4 \approx 0$, and has the solution,

$\tilde{\eta}_\perp \approx \sqrt{\frac{\sqrt{\beta^2 - 3a(1-\eta)} - \beta}{3(1-\eta)}}$, where $\eta = \frac{\tilde{\beta}_{15}^2 b^2}{4\lambda c}$. For $0 \leq |\tilde{\beta}_{15}b| \ll 1$ the solution corresponds to energy extrema

$$\tilde{g}_L^{ext} = \frac{-a^2(\beta^2 - 4(1-\eta)a)}{2(\beta^2 - 3(1-\eta)a)^{3/2} + 2\beta^3 - 9(1-\eta)\beta a} - \frac{\tilde{\beta}_{15}^2 b^2 \tilde{\eta}_\perp^4}{2(\lambda c + \tilde{\beta}_{13}b\tilde{\eta}_\perp^2)}$$ (B.10b)

In fact expression (B.10b) is nothing else but 2-nd order correction to the $G_{13}$-phase that is valid at $\lambda c > 0$.